\documentclass[final]{cvpr}

\usepackage{times}
\usepackage{epsfig}
\usepackage{graphicx}
\usepackage{amsmath}
\usepackage{amssymb}

\usepackage[accsupp]{axessibility}  

\usepackage[pagebackref=true,breaklinks=true,colorlinks=true,citecolor=teal,bookmarks=false]{hyperref}

\usepackage[table]{xcolor}
\usepackage{amssymb}
\usepackage{amsmath,amssymb,amsfonts}
\usepackage{graphicx}
\usepackage[draft]{fixme}
\usepackage{paralist}                           
\usepackage{bm}
\usepackage{float}
\usepackage{multirow}
\usepackage{booktabs}
\usepackage{nicefrac}
\usepackage{nth}
\usepackage[bf,small,skip=0pt]{caption}
\usepackage{subcaption}
\usepackage{wrapfig}
\usepackage{dsfont}
\usepackage{mathtools}
\usepackage{dsfont}
\usepackage{adjustbox}
\usepackage{multirow}
\usepackage{url}
\usepackage{color}
\usepackage{wrapfig}
\usepackage{mathrsfs}
\usepackage{ctable}
\usepackage{amsthm}
\usepackage[makeroom]{cancel}
\usepackage{tablefootnote}
\usepackage{afterpage}

\usepackage[linesnumbered,ruled,vlined]{algorithm2e}

\SetCommentSty{mycommfont}

\SetKwInput{KwInput}{Input}                
\SetKwInput{KwOutput}{Output}              

\usepackage{enumitem}
\setlist[itemize]{leftmargin=7.5mm}

\DeclareMathOperator*{\argmin}{arg\,min}

\newtheorem{theorem}{Theorem}[]
\newtheorem{lemma}{Lemma}[]

\newtheorem{definition}{Definition}[]

\numberwithin{equation}{section}

\newcommand{\supp}{Appx.}

\usepackage{pifont}

\newcommand{\cmark}{\ding{51}}%
\newcommand{\xmark}{\ding{55}}%

\definecolor{olive}{rgb}{0.42, 0.56, 0.14}

\def\cvprPaperID{4123} 
\def\confYear{CVPR 2022}

\begin{document}

\title{Aladdin: Joint Atlas Building and Diffeomorphic Registration Learning \\with Pairwise Alignment}

\author{Zhipeng Ding \quad Marc Niethammer\\
University of North Carolina at Chapel Hill, Chapel Hill, USA\\
{\tt\small \{zp-ding, mn\}@cs.unc.edu}
}

\maketitle

\begin{abstract}
Atlas building and image registration are important tasks for medical image analysis. Once one or multiple atlases from an image population have been constructed, commonly (1) images are warped into an atlas space to study intra-subject or inter-subject variations or (2) a possibly probabilistic atlas is warped into image space to assign anatomical labels. Atlas estimation and nonparametric transformations are computationally expensive as they usually require numerical optimization. Additionally, previous approaches for atlas building often define similarity measures between a fuzzy atlas and each individual image, which may cause alignment difficulties because a fuzzy atlas does not exhibit clear anatomical structures in contrast to the individual images. This work explores using a convolutional neural network (CNN) to jointly predict the atlas and a stationary velocity field (SVF) parameterization for diffeomorphic image registration with respect to the atlas. Our approach does not require affine pre-registrations and utilizes pairwise image alignment losses to increase registration accuracy.
We evaluate our model on 3D knee magnetic resonance images (MRI) from the OAI-ZIB dataset. Our results show that the proposed framework achieves better performance than other state-of-the-art image registration algorithms, allows for end-to-end training, and for fast inference at test time.\footnote{Source code: \url{https://github.com/uncbiag/Aladdin}.}
\end{abstract}

\section{Introduction}
\label{sec:intro}
Medical images can vary significantly across individuals due to different organ shapes and sizes. Therefore, representing medical images to account for such variability is crucial~\cite{joshi2004unbiased,evans2012brain}. A popular approach to represent and analyze medical images is through the use of one or more atlases~\cite{gee1993elastically,blezek2007atlas}. An atlas refers to a specific representation for a population of images; typically a form of generalized mean. An atlas can be used as a common coordinate system for the analysis of image segmentations or deformations or to track changes in longitudinal data, e.g., for the purpose of dose accumulation calculations in radiation treatment planning~\cite{zhang2017geometric,rigaud2019deformable}.
One simple way to obtain an atlas is by choosing one image from the image population~\cite{gee1993elastically,martin2008atlas}. However, this approach may introduce bias due to the particular selected image and in consequence may lead to inaccurate analysis results~\cite{joshi2004unbiased}. To obtain an unbiased atlas, many groupwise registration methods~\cite{joshi2004unbiased,lorenzen2005unbiased,wu2010groupwise-a} have been proposed. 
These approaches achieve more accurate overall alignments and more consistent registrations among the images by simultaneously estimating a population atlas. 

\begin{figure}[t!]
  \centering
  \includegraphics[width=\linewidth]{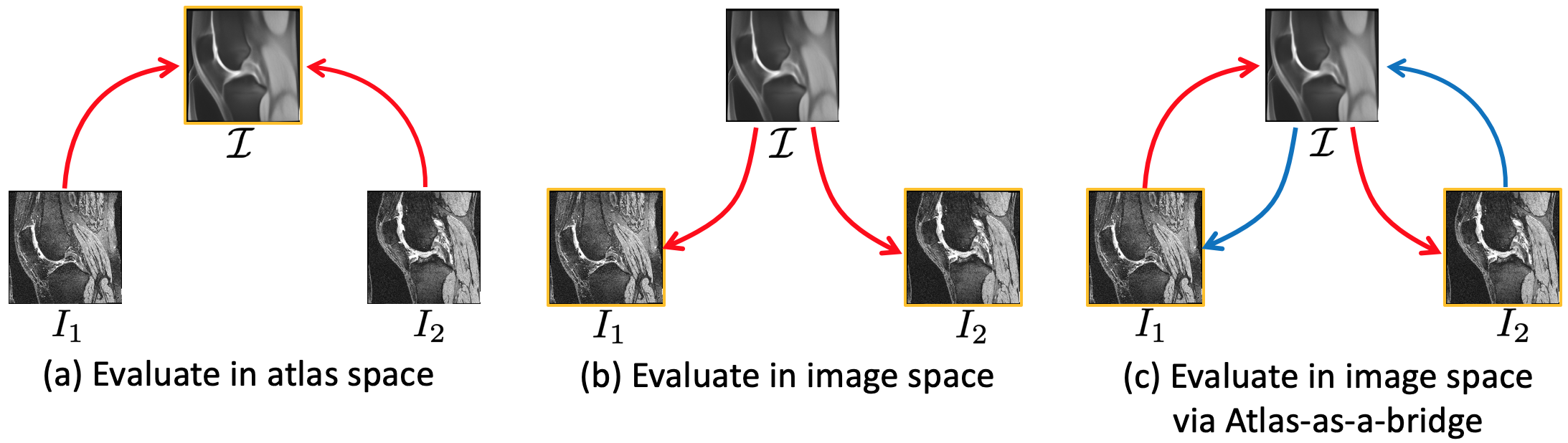}
  \caption[Atlas Evaluation]{Different atlas evaluation approaches with 2 images. (a) The labels of $I_1$ and $I_2$ are warped into atlas $\mathcal{I}$'s space, where their alignment is compared; (b) The probabilistic labels of atlas $\mathcal{I}$ are warped into image $I_1$'s (or $I_2$'s) space, where alignment is compared; (c) Using the atlas-as-a-bridge, each image's labels are warped to the other image's space, where alignment is compared.}
  \label{fig:atlas_evaluation}
  \vspace{-3mm}
\end{figure}

\begin{figure*}[th!]
  \centering
  \includegraphics[width=0.9\linewidth]{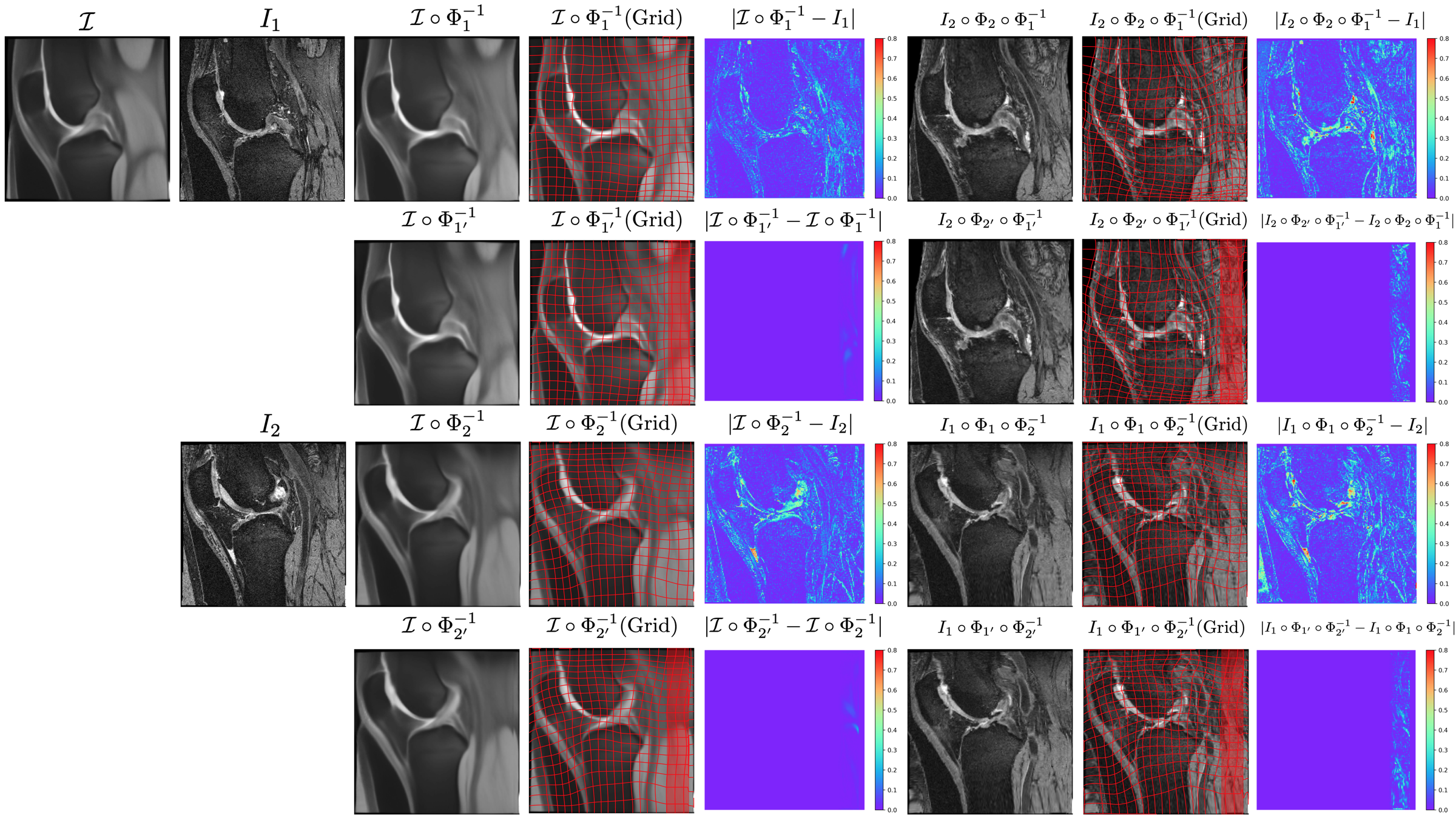}
  \caption[Similarity Measure Comparison]{Image similarity difference in a \emph{forward} atlas building model. \textbf{1st and 3rd row}: Left: atlas image $\mathcal{I}$ is warped to example image $I_1$ ($I_2$) via deformation map $\Phi_1^{-1}$ ($\Phi_2^{-1}$), and the absolute difference between the  warped atlas and the example image is evaluated. Right: two example images are warped into each other using the atlas as a bridge, and the absolute difference between them is evaluated. Image-to-image differences are more significant than atlas-to-image differences. \textbf{2nd and 4th row}: Left: a small perturbation on the deformation map (red highlighted rectangle area) is added, i.e. $\Phi_{1^{'}}^{-1} = \Phi_{1}^{-1} + \epsilon$, and the difference between the warped atlas image is evaluated (before/after perturbation). Right: example images are warped to each other using the atlas-as-a-bridge via the perturbed deformation maps, and their absolute difference is evaluated (before/after perturbation). Again, image-to-image differences are greater than atlas-to-image differences. Hence, an image-to-image similarity measure is expected to provide more alignment information than an atlas-to-image similarity measure because a fuzzy atlas does not exhibit the clear anatomic structures present in individual images.}
  \label{fig:pairwise_alignment}
  \vspace{-3mm}
\end{figure*}

The quality of an atlas is usually evaluated in combination with the quality of the image registration algorithm. For example, an atlas framework is often evaluated based on the sharpness or entropy of the atlas~\cite{wu2010groupwise-a,lorenzen2005unbiased}, the alignment of test images in the atlas space~\cite{he2020unsupervised} (Fig.~\ref{fig:atlas_evaluation}(a)), or the alignment of the warped atlas in test image space~\cite{dalca2019learning} (Fig.~\ref{fig:atlas_evaluation}(b)). These evaluation measures all have shortcomings. For sharpness, it is unclear if a sharper atlas is indeed better, as the atlas is usually used as a common coordinate space to which images are registered. Thus the ability of a registration algorithm to align corresponding points between images in atlas space may matter more than the sharpness of the atlas itself.
When measuring atlas quality via the alignment of test images in atlas space, the performance between different atlases is not directly comparable, because atlas quality is then measured in an \emph{atlas-specific} coordinate system\footnote{This could, of course, be avoided by weighting measures locally by the determinant of the Jacobian of the transformation map, but this would effectively amount to measuring differences in image-space.}. For example, using the same registration algorithm to warp images into a smaller atlas space will typically result in lower Dice scores than for a larger atlas space. Measuring atlas quality by alignment of the warped atlas in the test image space usually requires segmenting the atlas image itself. Such segmentations may not be accurate because an atlas may not clearly show anatomical structures (as it is a form of average image). To overcome these shortcomings of existing atlas evaluation measures, we propose to use the atlas as a bridge, where the atlas simply facilitates indirectly (i.e., by going through the atlas) warping image $I_1$ to image $I_2$'s space. Doing so we can now directly compare the alignment of the warped image $\tilde{I}_1$ to image $I_2$ (Fig.~\ref{fig:atlas_evaluation}(c)). The resulting alignment evaluation measure is (1) not directly affected by atlas variations, (2) does not require atlas segmentation, (3) and directly informs our proposed atlas-building approach. \emph{Hence, our atlas-as-a-bridge measure (Sec.~\ref{sec:performance_and_analysis}) is conceptually preferable to existing evaluation measures. We will use it to evaluate atlas-building and registration approaches in this work.}


Further, effectively using the atlas (via registrations) for analyses is as important as building the atlas in large-scale imaging studies, due to increasingly larger data sets, especially for MR images~\cite{ollier2005uk,ambellan2019automated}.
For example, to analyze the disease progression of 3D MR knee images, a first step is often to build a common atlas and then to warp all images to this atlas space. Hence, if there are many images to be analyzed, efficient registration algorithms that can avoid costly numerical optimization are desirable. Such efficient approaches, based on deep learning, have recently been proposed for registration and atlas building~\cite{dalca2019learning,he2020unsupervised,evan2020learning,sinclair2020atlas,dey2021generative}. However, these approaches use similarity measures between the fuzzy atlas and the anatomically more detailed images of the dataset. In consequence,  registration accuracy might be less accurate than directly registering images within the dataset. Fig.~\ref{fig:pairwise_alignment} demonstrates that an image-to-image similarity measure is more sensitive to anatomical structures than an atlas-to-image similarity measure. Therefore, \textbf{(H1)} \emph{we hypothesize that incorporating a pairwise image similarity loss is beneficial for alignment accuracy.} 

Furthermore, most existing atlas building methods~\cite{he2020unsupervised,he2021learning,dalca2019learning,evan2020learning} rely on affine pre-registrations to a chosen reference image~\cite{he2020unsupervised,he2021learning}. Alternatively, one can build an unbiased atlas based on affine transformations~\cite{joshi2004unbiased}.
To directly work with images which have not yet been affine aligned, Atlas-ISTN~\cite{sinclair2020atlas} simultaneously estimates a separate affine transformation before the nonparametric transformation and composes the two. Instead of separately considering affine and nonparametric transformations, we propose to predict a transformation which includes affine and nonparametric deformations. \textbf{(H2)} \emph{We hypothesize that our combined transformation prediction is as accurate as methods that treat affine and nonparametric transformations separately.}

This work proposes a framework (termed \emph{Aladdin}) that simultaneously builds an atlas for an image population and learns diffeomorphic registrations (to and from the atlas). 

Our \textbf{contributions} are as follows: (1) We develop an end-to-end joint atlas construction and diffeomorphic registration framework that, unlike existing approaches, does not require affine preregistration and that uses a pairwise image similarity losses to increase registration accuracy. (2) We provide a detailed mathematical explanation of the benefits of the pairwise alignment losses via the Euler-Lagrange equations of the optimization problem. (3) We comprehensively study evaluation measures for atlas building and registration. (4) We show that our proposed method can outperform previous learning-based atlas construction and registration approaches on a 3D MR knee dataset.
\section{Background and Related Work}
\label{sec:related_work}

There is a comprehensive literature on atlas construction~\cite{joshi2004unbiased,durrleman2008forward,he2020unsupervised,dalca2019learning}. Approaches can be divided into two different categories: \emph{backward} models and \emph{forward} models. The main difference between these two models lies in the warping directions.

\begin{definition}
Given a population of $N$ images $\{I_i\}_{i=1..N}$ that have been acquired by the same imaging modality, atlas building can be expressed as the following minimization problems:
\begin{align}
    \argmin_{\mathcal{I},\; \{\alpha_i\}} \sum_{i=1}^N \underbrace{\mathcal{L}_{sim}(\mathcal{I}, I_i \circ \Phi^{-1}_{\alpha_i}) + \lambda \mathcal{L}_{reg}(\Phi^{-1}_{\alpha_i})}_{\text{\normalsize backward model}}\,, \label{eq:backward_model}\\
    \argmin_{\mathcal{I},\; \{\beta_i\}} \sum_{i=1}^N \underbrace{\mathcal{L}_{sim}(\mathcal{I} \circ \Phi^{-1}_{\beta_i}, I_i) + \lambda \mathcal{L}_{reg}(\Phi^{-1}_{\beta_{i}})}_{\text{\normalsize forward model}}\,, \label{eq:forward_model}
\end{align}
where $\mathcal{L}_{sim}$ measures the dis-similarity between images, $\mathcal{L}_{reg}$ ensures transformation smoothness, $\mathcal{I}$ is the estimated atlas and  $\{\alpha_i\}$ (or $\{\beta_i\}$) parameterize the deformations, $\{\Phi^{-1}_{\cdot}\}$ (specified in the atlas space for the backward model and in the image space for the forward model). 
\end{definition}

These two models differ in the way they form the atlas. For simplicity,
we define $\mathcal{L}_{sim}$ as the squared $L_2$ norm, i.e., $\mathcal{L}_{sim}(I, J) = \|I-J\|^2_{2} = \langle I-J, I-J \rangle = \int_{\Omega} (I(x)-J(x))^2~dx$, where $\Omega$ is the image domain, $x\in \Omega$ is the position, and $\langle\cdot,\cdot\rangle$ is the usual $L_2$-product for square integrable vector-fields on $\Omega$. Denote the energy functional of Eq.~\eqref{eq:backward_model} as $E_1$ and Eq.~\eqref{eq:forward_model} as $E_2$. Assume the deformations $\{\Phi^{-1}_{\cdot}\}$ are diffeomorphic. By G\^{a}teaux variation w.r.t. $\mathcal{I}$ (see details in \supp~\ref{sec:variational_calculus}), we obtain the variations
\begin{align*}
    \delta E_1(\mathcal{I}; d\mathcal{I}) &= 2 \big\langle \sum_{i=1}^N \mathcal{I} - I_i \circ \Phi^{-1}_{\alpha_i}, d\mathcal{I}  \big\rangle \overset{!}{=} 0, \forall d\mathcal{I},\\
    \delta E_2(\mathcal{I}; d\mathcal{I}) &= 2 \big\langle \sum_{i=1}^N (\mathcal{I} - I_i \circ \Phi_{\beta_i})|D \Phi_{\beta_i}|, d\mathcal{I}  \big\rangle \overset{!}{=} 0, \forall d\mathcal{I}\,,  
\end{align*}
where $D$ denotes the Jacobian. If $\alpha_{i}$ and $\beta_{i}$ are fixed, we obtain the following optimal $\mathcal{I}^*$
\begin{align}
    (\text{\normalsize \emph{backward} model}) \quad& \mathcal{I}^* = \frac{1}{N} \sum_{i=1}^N I_i \circ \Phi^{-1}_{\alpha_i}\,, \label{eq:backward_opt_solution}\\
    (\text{\normalsize \emph{forward} model}) \quad& \mathcal{I}^* = \frac{\sum_{i=1}^N I_i \circ \Phi_{\beta_i} |D \Phi_{\beta_i}|}{\sum_{i=1}^N |D \Phi_{\beta_i}|}\,. \label{eq:forward_opt_solution}
\end{align}

In short, a \emph{backward} atlas will result in an average of all warped images while a \emph{forward} atlas will result in a weighted average of all warped images depending on the amount of deformation required.

\begin{figure*}[th!]
  \centering
  \includegraphics[width=1.0\linewidth]{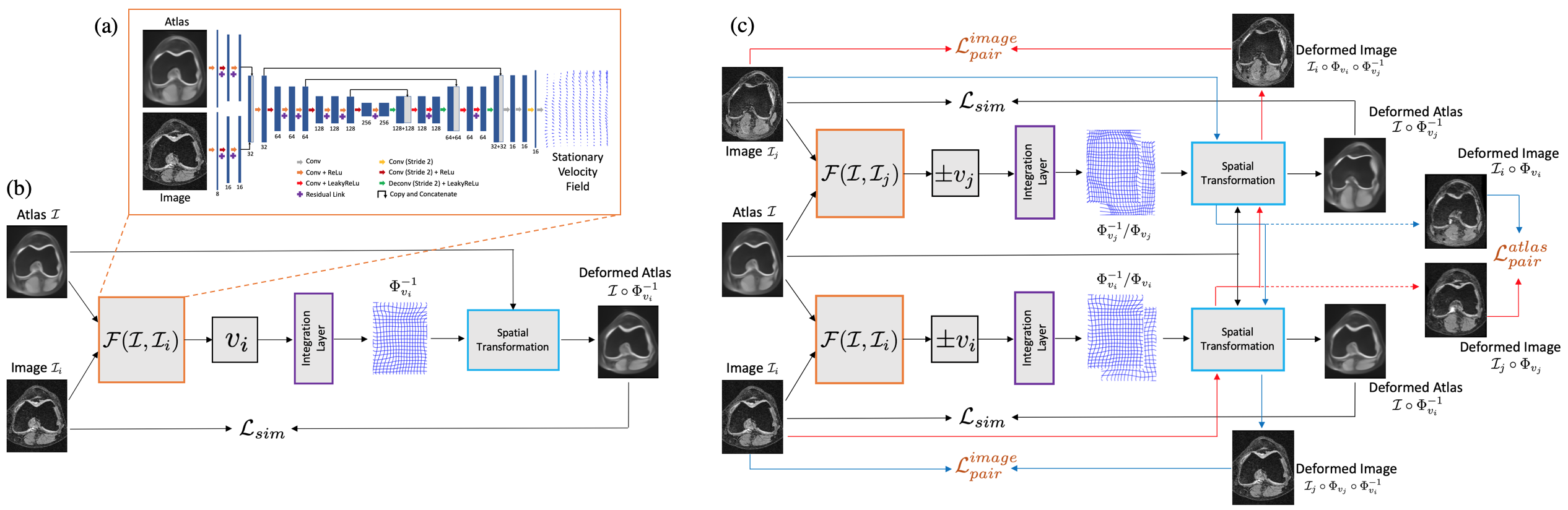}
  \caption[architecture]{(a) Architecture of the stationary velocity field (SVF) prediction network following a U-Net~\cite{cciccek20163d} structure. Feature dimensions (channels) of each convolutional layer are listed underneath each block. The network (denoted as $\mathcal{F}$) takes the atlas and an image as input, and outputs the stationary velocity field for the SVF model. (b) Standard \emph{forward} atlas building architecture. (c) Our proposed model with pairwise image alignment losses. Our model extends the standard model by incorporating pairwise image losses and by computing their alignments in atlas space ($\mathcal{L}^{atlas}_{pair}$) as well as in image space using the atlas as a bridge ($\mathcal{L}^{image}_{pair}$).}
  \label{fig:architecture_comparison}
  \vspace{-3mm}
\end{figure*}

In the following, we review the atlas building literature. Limitations are (1) the absence of a pairwise image alignment loss; (2) most approaches rely on affine pre-alignment; and (3) their evaluation measures are not comparable.

\textbf{Optimization-based \emph{backward} atlas building and registration models:} One of the most popular \emph{backward} atlas building methods was proposed by Joshi~\etal~\cite{joshi2004unbiased}, which iteratively warps all images to a tentative group mean image and averages all warped images to generate a new tentative group mean image, and then repeats these steps until convergence. Other approaches~\cite{bhatia2004consistent,lorenzen2005unbiased,avants2004geodesic,beg2003variational,lorenzen2006multi,bhatia2007groupwise,van2008encoding,fletcher2009geometric,wu2010groupwise-a,wu2010groupwise-b,wang2010groupwise,jia2010absorb,debroux2020variational} exist (see \supp~\ref{sec:additonal_related_work}).  \emph{These are optimization-based, i.e., their registrations are time consuming and hence inconvenient for large-scale analyses. Besides, these approaches assess the accuracy with respect to a specific atlas space which then no longer allows straightforward comparisons between approaches.} \textbf{Optimization-based \emph{forward} atlas building and registration models}~\cite{grenander1993general,ma2008bayesian,ma2010bayesian,durrleman2008forward,zhang2013bayesian,singh2013vector,zhang2015mixture} \emph{also suffer from large computational cost for registrations and require, just as backward approaches, affine pre-alignments} (see~\supp~\ref{sec:additonal_related_work}).

\textbf{Learning-based \emph{backward} atlas building and registration models:}
To overcome the large computational cost, He~\etal~\cite{he2020unsupervised} proposed a template-free unsupervised groupwise registration framework with multi-step refinements based on deep learning. Later work, uses a segmentation-assisted generative adversarial network~\cite{he2021learning} and an entropy loss to improve registration accuracy, transformation smoothness, and atlas unbiasedness. \emph{However, both methods rely on an affine pre-alignment, evaluate the registration performance on the variable atlas space, and do not use a pairwise alignment loss in image space.}


\textbf{Learning-based \emph{forward} atlas building and registration models:} Dalca~\etal~\cite{dalca2019learning} used a CNN to learn a conditional atlas to address heterogeneous groups of images. Yu~\etal~\cite{evan2020learning} proposed to learn a deformation map that warps a pre-specified atlas conditioned on subject-specific characteristics. Sinclair~\etal~\cite{sinclair2020atlas} proposed Atlas-ISTN, a deep-learning framework to jointly learn segmentation, registration and atlas construction. \emph{However, these approaches do not consider a pairwise image alignment loss to improve registration accuracy and their evaluation approaches require the annotation of the built atlas which is not a measurement that can be easily compared among approaches.}

In this work, we first show that pairwise image similarity losses improve registration accuracy. Second, we demonstrate how to learn a diffeomorphic transformation that combines affine and nonparametric transformations to avoid affine pre-alignment. Finally, we propose a novel way to evaluate the quality of an atlas and the predicted registrations. Tab.~\ref{tab:learning_atlas_building_comparisons} shows a comparison between our approach and other closely related learning-based approaches.

\begin{table}[ht]
\centering
\begin{adjustbox}{max width=1.0\linewidth}
\begin{tabular}{|c|c|c|c|}
\hline
Methods  & \begin{tabular}[c]{@{}c@{}}Handles\\ Affine Pre-alignment\end{tabular} & \begin{tabular}[c]{@{}c@{}}Incorporates\\ Pairwise Alignment\end{tabular} & \begin{tabular}[c]{@{}c@{}}Provides Reliable\\ Evaluation for Alignments\end{tabular} \\ \hline
He~\etal~\cite{he2020unsupervised} & \xmark& \cmark \tablefootnote{Note that this work only includes the pairwise alignment loss in atlas space, but not in image space.} & \xmark \\
\hline
He~\etal~\cite{he2021learning} & \xmark& \xmark & \xmark \\
\hline
Dalca~\etal~\cite{dalca2019learning} &\xmark& \xmark & \xmark \\ 
\hline
Yu~\etal~\cite{evan2020learning} &\xmark& \xmark & \xmark \\ 
\hline
Sinclair~\etal~\cite{sinclair2020atlas} &\cmark& \xmark & \xmark \\ 
\hline
\textit{Aladdin (Ours)} &\cmark& \cmark & \cmark \\
\hline
\end{tabular}
\end{adjustbox}
\caption{Comparison of learning-based atlas building approaches.}
\label{tab:learning_atlas_building_comparisons}
\vspace{-5mm}
\end{table}

\section{Methodology}
\label{sec:method}

Our proposed atlas building model is a \emph{forward} model because a \emph{forward} model evaluates the atlas-to-image similarity difference in a fixed image space (the spaces of the target images) while a \emph{backward} model evaluates in the evolving atlas space. Hence \textbf{(H3)} \emph{we hypothesize that a forward model is more accurate than a backward model}. 

\subsection{Pairwise Image Alignment}
Previous atlas building approaches evaluate image similarity between the atlas and each individual image. However, the built atlas image is usually fuzzy and does not show clear anatomical structures (e.g. see Fig.~\ref{fig:pairwise_alignment}), which can affect registration accuracy~\cite{wu2011sharpmean}. To increase registration accuracy, we propose using a pairwise alignment loss. Specifically, we propose to align individual images in atlas space \emph{and} image space to form the pairwise alignment loss.

Depending on how an atlas is used, image registrations may be performed  in the \emph{backward} and \emph{forward} directions. Registration in the \emph{backward} direction ensures that all individual images align well in the common atlas space. This is a useful property, e.g., when the atlas is used as a common space for population-level analyses. 
One way to encourage such a pairwise alignment is to measure the pairwise image similarity in \emph{atlas space}
\begin{equation}
\label{eq:pairwise_alignment_in_atlas_space}
    \mathcal{L}_{pair}^{atlas}(I_i, I_j) = \underbrace{\mathcal{L}_{pair}(I_i \circ \Phi_{\beta_i}, I_j \circ \Phi_{\beta_j})}_{I_i \text{ aligns with } I_j \text{ in atlas space}}\,.
\end{equation}


Registration in the \emph{forward} direction is to ensure that the atlas aligns well with each image in each individual image space. This is a useful property, e.g., when one wants to use atlas segmentations for the segmentation of a target image.
The related way to encourage pairwise alignment 
is to measure the pairwise image similarity in \emph{image space}
\begin{equation}
\begin{split}
\label{eq:pairwise_alignment_in_image_space}
     \mathcal{L}^{image}_{pair}(I_i, I_j) &= \underbrace{\mathcal{L}_{pair}(I_i \circ \Phi_{\beta_i} \circ \Phi^{-1}_{\beta_j}, I_j)}_{I_i \text{ aligns with } I_j \text{ in image space}} \\
     &+ \underbrace{\mathcal{L}_{pair}(I_j \circ \Phi_{\beta_j} \circ \Phi^{-1}_{\beta_i}, I_i)}_{I_j \text{ aligns with } I_i \text{ in image space}}\,.
\end{split}
\end{equation}

Fig.~\ref{fig:architecture_comparison}(c) illustrates both design choices; additional comparisons are in Sec.~\ref{sec:ele} and Sec.~\ref{sec:experiment}. 
A \emph{forward} atlas building model with pairwise image alignment can be formulated as 
\begin{align}
\label{eq:forward_model_with_paired_alignment}
     \argmin_{\mathcal{I},\; \{\beta_i\}} \sum_{(i, j) \in \Gamma}\Big[&\mathcal{L}_{sim}(\mathcal{I} \circ \Phi^{-1}_{\beta_i}, I_i) + \lambda \mathcal{L}_{reg}(\Phi^{-1}_{\beta_i}) \notag\\
    + &\mathcal{L}_{sim}(\mathcal{I} \circ \Phi^{-1}_{\beta_j}, I_j) + \lambda \mathcal{L}_{reg}(\Phi^{-1}_{\beta_j}) \\
    + &\gamma_1 \mathcal{L}^{atlas}_{pair}(I_i, I_j) + \gamma_2 \mathcal{L}^{image}_{pair}(I_i, I_j) \Big]\,, \notag
\end{align}
where $\lambda,\gamma_1,\gamma_2\geq 0$, $\Gamma = \{ (i, j) | i=1, 2, ..., N, j=1, 2, ..., N, i < j  \}$ is the set of all pairwise index combinations where the first index is smaller than the second index.
For $\gamma_1 = 0$ and $\gamma_2 = 0$ we obtain the vanilla \emph{forward} atlas building model. For $\gamma_1 >0$ and $\gamma_2 = 0$ we obtain the \emph{forward} model with pairwise alignment in atlas space. For $\gamma_1 = 0$ and $\gamma_2 > 0$ we obtain the \emph{forward} model with pairwise alignment in image space. Lastly, for $\gamma_1 >0$ and $\gamma_2 >0$ we obtain the \emph{forward} atlas building model with pairwise alignment in both atlas space and image space.

The expectation is that if all image pairs align well (through the transformations), in atlas space and image space, then atlas-to-image alignment is also accurate. Note that the new pairwise alignment loss terms do not explicitly change the optimal form of $\mathcal{I}^*$ \emph{given} a set of deformations because $\mathcal{L}^{atlas}_{pair}$ and $\mathcal{L}^{image}_{pair}$ do not contain $\mathcal{I}$ in their expressions and hence would not influence the G\^{a}teaux variation w.r.t. $\mathcal{I}$. The optimal form of $\mathcal{I}^*$ for given deformations remains as in Eq.~\eqref{eq:forward_opt_solution}. However, $\mathcal{L}^{atlas}_{pair}$ and $\mathcal{L}^{image}_{pair}$ influence the estimation of the optimal $\{\beta_i\}$ that parameterize the deformations $\{\Phi^{-1}_{\beta_i}\}$. Hence the atlas will get indirectly influenced. Sec.~\ref{sec:ele} provides a more detailed analysis.

\subsection{SVF based implementation}
\label{sec:svf_implementation}

We use a regularized stationary velocity field (SVF) parameterization~\cite{ashburner2007fast,hernandez2009registration,modat2012parametric,shen2019networks,dalca2019learning,sinclair2020atlas} for our registration model. 
In particular, we instantiate Eq.~\eqref{eq:forward_model_with_paired_alignment} as
\begin{align}
\label{eq:forward_svf_atlas_building_with_pairwise_alignment}
    \argmin_{\mathcal{I},\;\{v_i\}} &\sum_{(i, j) \in \Gamma} \Big [ \|I_i - \mathcal{I} \circ \Phi^{v_i}_{1, 0}\|^2_{2} + \lambda \sum_{k=1}^d \|\mathbf{H}_k(\Phi^{v_i}_{1, 0})\|_F^2 \notag\\
    &+ \|I_j - \mathcal{I} \circ \Phi^{v_j}_{1, 0}\|^2_{2} + \lambda \sum_{k=1}^d \|\mathbf{H}_k(\Phi^{v_j}_{1, 0})\|_F^2 \notag\\
    &+ \gamma_1\|I_i \circ \Phi^{v_i}_{0, 1} - I_j \circ \Phi^{v_j}_{0, 1}\|_2^2 \\
    + \gamma_2\Big(\|I_i &\circ \Phi^{v_i}_{0, 1} \circ \Phi^{v_j}_{1, 0} - I_j\|_2^2 + \|I_j \circ \Phi^{v_j}_{0, 1} \circ \Phi^{v_i}_{1, 0} - I_i\|_2^2\Big) \Big] \notag
\end{align}
where $v_i$ is the stationary field that parameterizes the deformation map $\Phi^{v_i}_t$ via $\frac{\partial \Phi^{v_i}_{t}}{\partial t} = v_i \circ \Phi^{v_i}_t$ over unit time $[0, 1]$, i.e., $\Phi^{v_i}_{t} (= \Phi^{v_i}_{0,t})$ is the integration of the above ODE from time 0 to time t, with $\Phi^{v_i}_0 = \text{Id}$ which is the identity map. In our case, we obtain the final transformation map $\Phi^{v_i}_{0, 1}$ and its inverse $\Phi^{v_i}_{1, 0}$ by integrating the stationary fields $v_i$ and $-v_i$ over $t = [0, 1]$ through \emph{scaling and squaring}~\cite{arsigny2006log,dalca2019learning,sinclair2020atlas} (see \cite{dalca2019unsupervised} for detailed explanations). $\Phi^{v_i}_{s,t}$ is the composition $\Phi^{v_i}_{s,t} = \Phi^{v_i}_{t} \circ (\Phi^{v_i}_{s})^{-1}$. 
$\mathbf{H}_k(\Phi^{v_i}_{1, 0})$ is the Hessian matrix of the $k$-th component of $\Phi^{v_i}_{1, 0}$, and $d$ denotes the dimension ($d=3$ in our case). The squared Frobenius norm of the Hessian of the components of the transformation map is used as a loss to obtain a regularized (i.e., sufficiently smooth) velocity field. Other regularizers, e.g., regularizers directly on the velocity field, are, of course, possible.


\noindent\textbf{Deep Learning Model.} We use a deep CNN (denoted as $\mathcal{F}$ in Fig.~\ref{fig:architecture_comparison}(a)) to predict the stationary velocity field $v_i$ in our SVF based framework (Fig.~\ref{fig:architecture_comparison}(c)). The overall loss for one pair of images is defined as in the $[\cdot]$ of Eq.~\eqref{eq:forward_svf_atlas_building_with_pairwise_alignment}. There are two ways to obtain the atlas in our framework: using Eq.~\eqref{eq:forward_opt_solution} or learning the atlas. Note that Eq.~\eqref{eq:forward_opt_solution} only applies when we use mean squared error (MSE) as the similarity measure. Other more advanced similarity measures (e.g. normalized cross correlation (NCC)) do not lead to a closed-form solution. For our experiments, we explore using Eq.~\eqref{eq:forward_opt_solution} when using MSE. We use alternating optimization: (1) Keeping the atlas fixed, we update the parameters of $\mathcal{F}$ over every 10 epochs; (2) After every 10 epochs, we then update the atlas via Eq.~\eqref{eq:forward_opt_solution} with the latest warped training images. When the atlas image itself is also learnable, we use another optimization strategy to update the atlas. Different from updating the parameters of $\mathcal{F}$ in each iteration, 
we accumulate the atlas image gradient for each batch and then update the atlas image using the accumulated gradient at the end of each epoch. In this way, the gradient with respect to the atlas image takes into account all training images and will not be distracted by extreme data points in a small batch.
For both atlas-building strategies, the initial atlas is estimated as the mean of all training images.

\noindent\textbf{Affine Pre-registration.} Different from other atlas building approaches, we use a bending energy~\cite{rueckert1999nonrigid,xu2019deepatlas,ding2021votenet++} to regularize transformation maps. Hence, affine pre-registration is not necessary as the regularizer penalizes second derivatives (which zeros out any affine contributions)\footnote{The same reasoning also holds for curvature registration models~\cite{fischer2003curvature}, which therefore also do not require affine pre-registrations.}. Consequently, our framework simultaneously captures affine and nonparametric deformations. Details can be found in \supp~\ref{app:regularization_more_explanation}.

\begin{figure*}[t!]
  \centering
  \includegraphics[width=\linewidth]{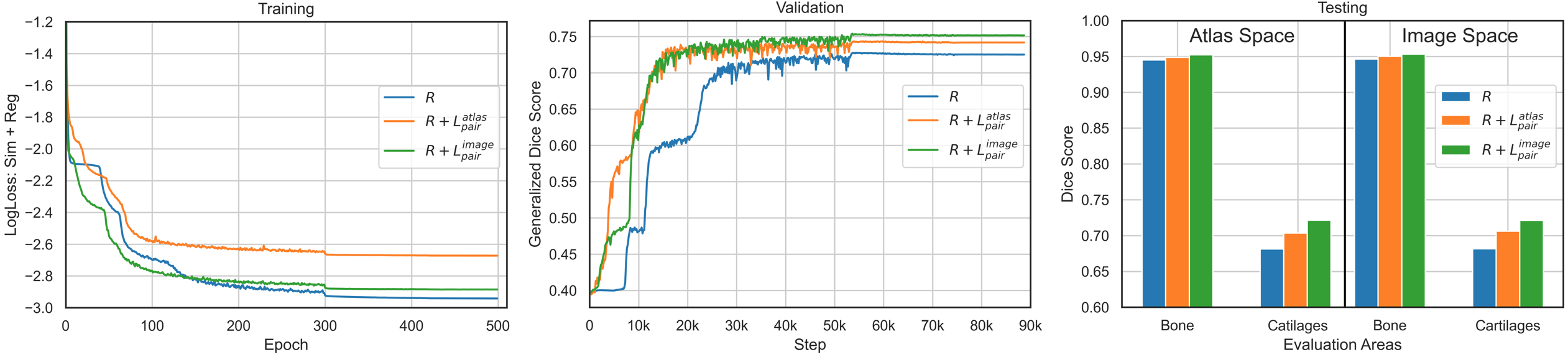}
  \caption{Comparison of a vanilla atlas-to-image registration model (denoted as $R$) and its improvements using pairwise alignment (i.e. Eq.~\eqref{eq:forward_svf_atlas_building_with_pairwise_alignment}, denoted as $R + \mathcal{L}^{atlas}_{pair}$ and $R + \mathcal{L}_{pair}^{image}$) for a fixed atlas. \textbf{Left:} average training loss curves. We only compare the similarity measure $\mathcal{L}_{sim}$ and the regularization term $\mathcal{L}_{reg}$, making the three methods comparable. $R$ has the lowest loss, both $\mathcal{L}^{atlas}_{pair}$ and $\mathcal{L}^{image}_{pair}$ help prevent overfitting, and the latter works better by continuing to push the loss down. \textbf{Middle:} validation results. We notice that both pairwise alignment terms ($\mathcal{L}_{pair}^{image}$ and $\mathcal{L}_{pair}^{atlas}$) speed-up the convergence and increase the overall performance. Further, $\mathcal{L}_{pair}^{image}$ works better than $\mathcal{L}_{pair}^{atlas}$. \textbf{Right:} testing alignment performance in atlas space and image space (These two evaluation measures are reliable in this case because the atlas is fixed). Both $\mathcal{L}^{atlas}_{pair}$ and $\mathcal{L}^{image}_{pair}$ terms improve the performance, and $\mathcal{L}^{image}_{pair}$ performs better than $\mathcal{L}^{atlas}_{pair}$. 
  } 
  \label{fig:pairwise_loss_demo}
  \vspace{-3mm}
\end{figure*}

\subsection{Euler-Lagrange Equations}
\label{sec:ele}

To illustrate the benefit of the pairwise alignment terms, we fix atlas $\mathcal{I}$ in Eq.~\eqref{eq:forward_svf_atlas_building_with_pairwise_alignment} and calculate the Euler-Lagrange equations with respect to $v_i$.

\noindent\textbf{Theorem~\ref{thm:ele_atlas_atlas_image_together}.} 
\textit{Given a continuous differentiable idealized atlas image $\mathcal{I}$ and a population of noisy observed anatomies $I_i$ ($i = 1, ..., N$), the \{$v_i^*$\} minimizing the energy functional
\begin{align}
\label{eq:pairwise_atlas_building}
    E_1(\{v_i\}) = &\sum_{(i, j) \in \Gamma} \Big[ \lambda \text{Reg}(v_i) + \| \mathcal{I} \circ \Phi_{1, 0}^{v_i} - I_i \|^2_{2} \notag\\
    &+ \lambda \text{Reg}(v_j) + \| \mathcal{I} \circ \Phi_{1, 0}^{v_j} - I_j \|^2_{2} \notag\\
    &+ \gamma_1 \| I_i \circ \Phi_{0,1}^{v_i} - I_j \circ \Phi_{0,1}^{v_j} \|^2_{2} \\
    + \gamma_2 \Big( \| I_i \circ \Phi_{0,1}^{v_i} &\circ \Phi_{1,0}^{v_j} - I_j \|^2_{2} + \| I_j \circ \Phi_{0,1}^{v_j} \circ \Phi_{1,0}^{v_i} - I_i \|^2_{2} \Big)\Big] \notag
\end{align}
satisfy the Euler-Lagrange equation
\begin{align}
\label{eq:ele_pairwise_atlas_image_together}
    &(N-1)\Big(\lambda \nabla_{v_i^*}\text{Reg}(v_i^*) -  2\int_0^1 \Big( |D\Phi_{t,1}^{v_i^*}|  (\mathcal{J}_t - J^i_t) \nabla \mathcal{J}_t \Big)~dt\Big) \notag\\
    &- 2N\gamma_1 \int_0^1 \big( |D\Phi_{t,0}^{v_i^*}| (\hat{\mathcal{J}}_t - J^i_t) \nabla J^i_t \big)~dt \\
    &- 2\gamma_2 \int_0^1 \Big( \big( \sum_{j=1}^N |D\Phi_{0,1}^{v^*_j}| \big)|D\Phi_{t,0}^{v^*_i}| (\tilde{\mathcal{J}}_t - J^i_t) \nabla J^i_t \Big)~dt \notag\\
    &- 2\gamma_2 \int_0^1 \sum_{j=1}^N |D\Phi_{t,1}^{v^*_i}| (J_t^{j, i} - J^i_t) \nabla J_t^{j, i}~dt = 0, \forall i \notag
\end{align}
where $\mathcal{J}_t \overset{.}{=} \mathcal{I} \circ \Phi^{v_i^*}_{t, 0}$, $J^i_t \overset{.}{=} I_i \circ \Phi^{v_i^*}_{t, 1}$, $\hat{\mathcal{J}}_t \overset{.}{=} \frac{\sum_{j=1}^N I_i \circ \Phi_{0,1}^{v_j^*}}{N} \circ \Phi_{t,0}^{v_i^*}$, $J_t^{j, i} \overset{.}{=} I_j \circ \Phi_{0,1}^{v^*_j} \circ \Phi_{t,0}^{v^*_i}$, $\tilde{\mathcal{J}}_t \overset{.}{=} \frac{\sum_{j=1}^N |D\Phi_{0,1}^{v^*_j}| I_j \circ \Phi_{0,1}^{v^*_j}}{\sum_{j=1}^N |D\Phi_{0,1}^{v^*_j}|} \circ \Phi_{t,0}^{v^*_i}$.}

For both optimization approaches or deep learning approaches, we need to iteratively update $v_i$. Without loss of generality, we use the standard steepest descent scheme
\begin{equation}
\label{eq:steepest_descent}
    v_i^{k+1} = v_i^k - \epsilon \nabla_{v_i} E
\end{equation}
to explain the benefit of the pairwise terms. Specifically, at optimality the Euler-Lagrange equations need to hold. This means the left-hand side of Eq.~\ref{eq:ele_pairwise_atlas_image_together} needs to be equal to zero. Away from optimality this left-hand side corresponds to the energy-gradient\footnote{To draw an analogy: in finite dimensions, fulfilling that the gradient of an objective function is zero at optimality corresponds to the Euler-Lagrange (EL) equation. Hence, evaluating the gradient (at optimality or not) corresponds to the part of the EL equation that needs to be equal to 0.} with respect to $v_i$, i.e., $\nabla_{v_i} E$. 
Therefore, adding pairwise terms will speed-up the convergence of $v_i$ in Eq.~\eqref{eq:steepest_descent}. Furthermore, $\mathcal{L}_{pair}^{atlas}$ encourages atlas similarity to the \emph{backward} atlas (Eq.~\eqref{eq:backward_opt_solution}) as indicated by $\hat{\mathcal{J}}_{t}$ thus helping image alignment in atlas space; $\mathcal{L}_{pair}^{image}$ encourages atlas similarity to the \emph{forward} atlas (Eq.~\eqref{eq:forward_opt_solution}) as indicated by $\tilde{\mathcal{J}}_{t}$ thus helping image alignment in image space. Hence $\mathcal{L}^{image}_{pair}$ is more suitable in our model.

To verify our observation above, we design a simple experiment to compare a vanilla atlas-to-image registration model to show improvements obtained by the pairwise loss terms. Fig.~\ref{fig:pairwise_loss_demo} shows the results. We can see that the pairwise terms can indeed speed-up the convergence for training and increase the accuracy for testing.

\section{Experimental Results and Discussion}
\label{sec:experiment}
\noindent
\textbf{Dataset.} We use a 3D knee MRI dataset (507 images) from the Osteoarthritis Initiative (OAI) for our experiments. This dataset includes manual segmentations for four structures (femur, tibia, femoral cartilage and tibial cartilage) for all images~\cite{ambellan2019automated}. All images are of size $384\times384\times160$, and each voxel is of size $0.36\times0.36\times0.7 mm^3$. We normalize the intensities of each image using an affine intensity transformation such that the $0.1th$ and the $99.9th$ percentile are mapped to $0$, $1$ and clamp values smaller than $0$ and larger than $1$ to avoid outliers. All images and manual segmentations are downsampled to size $192\times192\times80$. Our Train/Validate/Test split is $354/53/100$.

\noindent
\textbf{Training and Hyperparameters.} Our model is implemented with PyTorch. We train using ADAM~\cite{kingma2014adam} over 500 epochs. The learning rate is $10^{-4}$ with a batch size of 2. The hyperparameters of the model have an influence on the sharpness of the atlas, the smoothness of the deformation fields, and the accuracy of the registrations. In all experiments, we fine-tune hyperparameters based on a hold-out validation dataset. Details can be found in \supp~\ref{sec:details_and_hyperparameters}.

\noindent
\textbf{Evaluation Criteria.} As discussed in Sec.~\ref{sec:intro}, previous evaluation measures in atlas space and image space are not reliable because they are affected by atlas variations or the accuracy of the atlas segmentation. We propose to use the atlas as a bridge to evaluate alignments in each image's space. 
To evaluate foldings of the transformation maps from the predicted registrations, we use the determinant of the Jacobian of these maps, i.e., $J_{\Phi}(x) \overset{.}{=}|D\Phi^{-1}(x)|$, and count folds (defined as $|\{ x: J_{\Phi}(x) < 0 \}|$) in each image.

\noindent
\textbf{Baselines.} First, to illustrate the benefit of the pairwise terms, we compare the atlas building approaches before and after adding the pairwise similarity terms in two models: a previously proposed \emph{forward} model~\cite{dalca2019learning} by Dalca~\etal and our proposed model (\emph{Aladdin}). We also compare with standard atlas building methods\footnote{We only use a subset of 60 images from the training dataset to build the atlas for these two optimization-based methods, because of the required runtime. Using all training images would take several days of computation.}:  the method by Joshi~\etal~\cite{joshi2004unbiased} and ABSORB~\cite{jia2010absorb}. Furthermore, we use a deep registration model (Voxelmorph)~\cite{balakrishnan2019voxelmorph} to replace costly optimization-based registration in a \emph{backward} model~\cite{joshi2004unbiased} setup and a \emph{forward} model setup with and without affine pre-registration. A learning-based \emph{backward} model~\cite{he2020unsupervised}\footnote{This method does not generate an atlas, but we add the atlas according to Eq.~\eqref{eq:backward_opt_solution} in the implementation because we found that without an atlas the registration performance is significantly worse than with an atlas.} is also included. Finally, we include results without registration (denoted as none) and optimization-based affine pre-registration (via Nifty-Reg~\cite{rueckert1999nonrigid,modat2010fast,modat2014global,ourselin2001reconstructing}) results as baselines. For our approach and \cite{dalca2019learning}, the inverse deformation calculation is available. For all other baseline models, where inverse deformation maps are not available as part of the implementations, we obtain them numerically by solving $\argmin_{\phi}||\Phi \circ \phi - \text{Id}||_2^2 + ||\phi \circ \Phi - \text{Id}||_2^2$, where $\Phi$ is the known deformation map and $\phi$ is the sought-for inverse deformation map. See details in \supp~\ref{sec:details_and_hyperparameters}.

\subsection{Performance and Analysis}
\label{sec:performance_and_analysis}
\noindent\textbf{Comparison between evaluation measures.}
Let $S^k_1$, $S^k_2$ be the segmented voxels of structure $k$ for image $I_1$, $I_2$, respectively. We quantify the volume overlap for structure $k$ using the Dice score\cite{dice1945measures}, i.e., $\text{Dice}(S^k_1, S^k_2) = 2 \cdot \frac{|S^k_1 \cap S^k_2|}{|S^k_1| + |S^k_2|}$.
A Dice score of 1 indicates that the anatomy matches perfectly, and a score of 0 indicates that there is no overlap.
Further, assume we have the deformations $\{\Phi_i\}$ that warp $M$ test images $\{I_i\}$ and segmentation $\{S_i\}$ to atlas $\mathcal{I}$, and the corresponding inverse deformations $\{\Phi^{-1}_i\}$. Let $\mathcal{S}$ be the estimated segmentation of $\mathcal{I}$ (e.g. by automatic or manual labeling). 
Let $\texttt{V}(\{\cdot\})$ be the plurality voting scheme to obtain the consensus segmentation among multiple segmentations. We define the following evaluation measures
\begin{align*}
    d^{atlas}_k &= \frac{1}{M}\sum_{i=1}^M \text{Dice}\big(S^k_i \circ \Phi_i \,, \texttt{V}(\{S^k_i \circ \Phi_i\}_{i=1...M})\big)\,,\\
    d^{image}_k &= \frac{1}{M}\sum_{i=1}^M \text{Dice}\big(\mathcal{S} \circ \Phi^{-1}_i \,, S^k_i\big)\,,\\
    d^{bridge}_k &= \frac{1}{M}\sum_{j=1}^M \text{Dice}\big(\texttt{V}(\{(S^k_i \circ \Phi_i) \circ \Phi^{-1}_j\}_{i=1...M}^{i \neq j}) \,, S^k_j \big)\,.
\end{align*}
where for structure $k$, $d^{atlas}_k$ is the atlas space  (Fig.~\ref{fig:atlas_evaluation}(a)), $d^{image}_k$ the image space  (Fig.~\ref{fig:atlas_evaluation}(b)), and $d^{bridge}_k$ is the atlas-as-a-bridge (Fig.~\ref{fig:atlas_evaluation}(c)) evaluation measure. 
For $d^{atlas}_k$, both atlas variations and registration errors contribute. For $d^{image}_{k}$, both estimated atlas segmentation errors and registration errors contribute. 
For $d^{bridge}_k$, only registration errors contribute. Hence, \emph{$d^{bridge}_k$ is  preferable}.

\setlength\tabcolsep{12pt}
\begin{table*}[t!]
\centering
\begin{adjustbox}{max width=0.8\textwidth}
\begin{tabular}{ccccccccccc}
\specialrule{.15em}{.05em}{.05em}
\multicolumn{1}{c}{\multirow{3}{*}{Index}}&\multicolumn{1}{c}{\multirow{3}{*}{Models}} &\multicolumn{1}{c}{\multirow{3}{*}{\shortstack{Affine \\Pre-alignment}}} &\multicolumn{1}{c}{\multirow{3}{*}{\shortstack{Atlas \\Acquisition}}} &\multicolumn{1}{c}{\multirow{3}{*}{\shortstack{Similarity \\Measure}}}& \multicolumn{2}{c}{\multirow{3}{*}{\shortstack{Pairwise \\Alignment Losses}}} & \multicolumn{3}{c}{Volume Dice (\%) $\uparrow$}&\multirow{3}{*}{Folds}\\ 
\cmidrule(lr){8-10}
&&&&&&& \multicolumn{3}{c}{Atlas-as-a-bridge}&\\
\cmidrule(lr){8-10}
\cmidrule(lr){6-7}
&&&&&$\mathcal{L}^{atlas}_{pair}$&$\mathcal{L}^{image}_{pair}$& \textit{Bone}  & \textit{Cartilages} & \textit{All} &\\
\specialrule{.15em}{.05em}{.05em}
A&\multicolumn{1}{c}{\multirow{1}{*}{None}}&\xmark&Eq.~\eqref{eq:backward_opt_solution}&--&--&--&76.70(6.65)&0.00(0.00)&38.37(3.25)&0.0\\
\hline
B&\multicolumn{1}{c}{\multirow{1}{*}{Affine}}&\xmark&Eq.~\eqref{eq:backward_opt_solution}&MSE&--&--&90.35(1.89)&38.44(9.35)&64.39(4.98)&0.0\\
\hline
C&\multicolumn{1}{c}{\multirow{1}{*}{Joshi~\etal~\cite{joshi2004unbiased}}}&\cmark&\multicolumn{1}{c}{\multirow{1}{*}{Eq.~\eqref{eq:backward_opt_solution}}}&\multicolumn{1}{c}{\multirow{1}{*}{MSE}}&\multicolumn{1}{c}{\multirow{1}{*}{--}}&\multicolumn{1}{c}{\multirow{1}{*}{--}}&94.98(1.83)&73.58(4.47)&84.28(2.85)&24.39\\
\hline
\multicolumn{1}{c}{\multirow{1}{*}{D}}&\multicolumn{1}{c}{\multirow{1}{*}{ABSORB~\cite{jia2010absorb}}}&\multicolumn{1}{c}{\multirow{1}{*}{\cmark}}&\multicolumn{1}{c}{\multirow{1}{*}{\shortstack{ Hierarchical}}}&\multicolumn{1}{c}{\multirow{1}{*}{MSE}}&\multicolumn{1}{c}{\multirow{1}{*}{--}}&\multicolumn{1}{c}{\multirow{1}{*}{--}}&\multicolumn{1}{c}{\multirow{1}{*}{95.18(1.73)}}&\multicolumn{1}{c}{\multirow{1}{*}{73.84(4.24)}}&\multicolumn{1}{c}{\multirow{1}{*}{84.51(2.59)}}&\multicolumn{1}{c}{\multirow{1}{*}{37.49}}\\
\hline
\multicolumn{1}{c}{\multirow{2}{*}{E}}&\multicolumn{1}{c}{\multirow{2}{*}{Voxelmorph~\cite{balakrishnan2019voxelmorph}}}&\xmark&\multicolumn{1}{c}{\multirow{2}{*}{Eq.~\eqref{eq:backward_opt_solution}}}&\multicolumn{1}{c}{\multirow{2}{*}{MSE}}&\multicolumn{1}{c}{\multirow{2}{*}{--}}&\multicolumn{1}{c}{\multirow{2}{*}{--}}&91.16(3.06)&59.13(10.42)&75.15(6.49)&237.04\\
&&\cmark&&&&&93.04(1.48)&73.34(4.18)&83.19(2.44)&18.76\\
\hline
\multicolumn{1}{c}{\multirow{2}{*}{F}}&\multicolumn{1}{c}{\multirow{2}{*}{Voxelmorph~\cite{balakrishnan2019voxelmorph}}}&\xmark&\multicolumn{1}{c}{\multirow{2}{*}{Eq.~\eqref{eq:forward_opt_solution}}}&\multicolumn{1}{c}{\multirow{2}{*}{MSE}}&\multicolumn{1}{c}{\multirow{2}{*}{--}}&\multicolumn{1}{c}{\multirow{2}{*}{--}}&91.47(3.03)&59.99(10.43)&75.73(6.48)&272.22\\
&&\cmark&&&&&93.09(1.47)&73.37(4.14)&83.23(2.41)&15.47\\
\hline
\multicolumn{1}{c}{\multirow{2}{*}{G}}&\multicolumn{1}{c}{\multirow{2}{*}{He~\etal~\cite{he2020unsupervised}}}&\xmark&\multicolumn{1}{c}{\multirow{2}{*}{Eq.~\eqref{eq:backward_opt_solution}}}&\multicolumn{1}{c}{\multirow{2}{*}{NCC}}&\multicolumn{1}{c}{\multirow{2}{*}{\cmark}}&\multicolumn{1}{c}{\multirow{2}{*}{--}}&91.91(2.62)&60.06(9.71)&75.99(6.61)&482.82\\
&&\cmark&&&&&95.13(1.32)&72.03(6.55)&83.58(3.53)&476.45\\
\hline
\multicolumn{1}{c}{\multirow{4}{*}{H}}&\multicolumn{1}{c}{\multirow{4}{*}{Dalca~\etal~\cite{dalca2019learning}}}&\multicolumn{1}{c}{\multirow{4}{*}{\xmark}}&\multicolumn{1}{c}{\multirow{4}{*}{Learning}}&\multicolumn{1}{c}{\multirow{4}{*}{MSE}}&\xmark&\xmark&93.65(3.31)&60.82(9.75)&77.24(6.28)&34.99\\
&&&&&\cmark&\xmark&93.95(2.65)&61.44(7.97)&77.70(5.06)&51.76\\
&&&&&\xmark&\cmark&95.07(1.96)&63.27(5.81)&79.17(3.61)&96.76\\
&&&&&\cmark&\cmark&94.53(2.13)&61.55(6.82)&78.04(4.21)&116.30\\
\hline
\multicolumn{1}{c}{\multirow{4}{*}{I}}&\multicolumn{1}{c}{\multirow{4}{*}{Dalca~\etal~\cite{dalca2019learning}}}&\multicolumn{1}{c}{\multirow{4}{*}{\cmark}}&\multicolumn{1}{c}{\multirow{4}{*}{Learning}}&\multicolumn{1}{c}{\multirow{4}{*}{MSE}}&\xmark&\xmark&95.29(1.29)&72.48(6.34)&83.89(3.69)&78.97\\
&&&&&\cmark&\xmark&95.69(1.24)&73.54(6.24)&84.62(3.61)&131.20\\
&&&&&\xmark&\cmark&95.72(1.23)&74.18(5.98)&84.95(3.48)&86.35\\
&&&&&\cmark&\cmark&95.71(1.24)&72.47(6.44)&84.09(3.71)&133.55\\
\hline
\multicolumn{1}{c}{\multirow{4}{*}{J}}&\multicolumn{1}{c}{\multirow{4}{*}{\emph{Aladdin}}}&\multicolumn{1}{c}{\multirow{4}{*}{\xmark}}&\multicolumn{1}{c}{\multirow{4}{*}{Eq.~\eqref{eq:forward_opt_solution}}}&\multicolumn{1}{c}{\multirow{4}{*}{MSE}}&\xmark&\xmark&95.81(1.16)&73.97(5.95)&84.89(3.43)&14.54\\
&&&&&\cmark&\xmark&95.83(1.16)&74.16(5.75)&84.99(3.41)&36.82\\
&&&&&\xmark&\cmark&\textbf{96.09(1.03)}&\textbf{74.86(5.43)}&\textbf{85.48(3.09)}&7.95\\
&&&&&\cmark&\cmark&95.71(1.23)&73.74(6.15)&84.72(3.54)&26.00\\
\hline
\multicolumn{1}{c}{\multirow{4}{*}{K}}&\multicolumn{1}{c}{\multirow{4}{*}{\emph{Aladdin}}}&\multicolumn{1}{c}{\multirow{4}{*}{\xmark}}&\multicolumn{1}{c}{\multirow{4}{*}{Learning}}&\multicolumn{1}{c}{\multirow{2}{*}{MSE}}&\xmark&\xmark&95.77(1.22)&73.06(6.46)&84.41(3.71)&18.86\\
&&&&&\xmark&\cmark&96.08(1.17)&74.39(5.83)&85.24(3.38)&6.74\\
\cline{5-11}
&&&&\multicolumn{1}{c}{\multirow{2}{*}{NCC}}&\xmark&\xmark&96.03(1.06)&73.86(6.28)&84.94(3.55)&53.03\\
&&&&&\xmark&\cmark&\textbf{96.27(0.93)}&\textbf{75.11(5.17)}&\textbf{85.69(2.92)}&31.61\\
\specialrule{.15em}{.05em}{.05em}
\end{tabular}
\end{adjustbox}
\caption{Evaluation of OAI atlas building performance. Mean (std) results are reported in two \emph{Bone} areas, two \emph{Cartilage} areas, and for \emph{All} areas together. Our proposed model (\emph{Aladdin}) outperforms other atlas building approaches. Best two results are bolded.}
\label{tab:atlas_metrics}
\vspace{-4mm}
\end{table*}

\noindent\textbf{Comparison between atlas building approaches.}
Tab.~\ref{tab:atlas_metrics} shows the alignment performance for all baseline methods and our proposed approach (\emph{Aladdin}).  \emph{Aladdin} outperforms other atlas building and registration approaches in terms of the atlas-as-a-bridge evaluation measure in both the \emph{bone} and \emph{cartilage} areas. It also shows fewer folds. To compare fairly, we choose the same similarity measure (MSE) for most methods except in row G where we use NCC in \cite{he2020unsupervised} and in row K where we check whether it is beneficial to use more advanced similarity measures with \emph{Aladdin}.
All hyperparameters are fine-tuned based on the same validation dataset to present the best possible performance.

\noindent\textbf{Q1$\mapsto$H1: Are pairwise image alignment losses helpful?} Pairwise alignment losses are designed to increase registration accuracy as image-to-image similarity measures provide richer anatomical information than atlas-to-image similarity measures. Rows H, I, J of Tab.~\ref{tab:atlas_metrics} show that \emph{both $\mathcal{L}^{atlas}_{pair}$ and $\mathcal{L}^{image}_{pair}$ help to increase accuracy, and $\mathcal{L}^{image}_{pair}$ outperforms $\mathcal{L}^{atlas}_{pair}$.} Interestingly, combining both pair-wise losses reduces accuracy over only using $\mathcal{L}^{image}_{pair}$. This can partly be explained by the fact that $\mathcal{L}^{image}_{pair}$ and $\mathcal{L}^{atlas}_{pair}$ resort to different mean images
as indicated in Sec.~\ref{sec:ele}. 

\noindent\textbf{Q2$\mapsto$H2: Is affine pre-registration a necessity?} The baseline methods in Tab.~\ref{tab:atlas_metrics} require affine pre-registrations. Consider the comparisons in the E, F, G, and H-I rows: under the same condition, a model with affine pre-registration works much better than one without. This is particulary obvious for the cartilage segmentations, which are small and thin. Thus, properly handling affine pre-registration is critical for atlas building methods. Our approach (row J), achieves better performance than the other methods without using affine pre-registration. This is because our regularization term allows the model to capture both affine and nonparametric transformations. \emph{Therefore, affine pre-registration is indeed not necessary in our framework.}

\noindent\textbf{Q3$\mapsto$H3: Should we use a \emph{forward} or \emph{backward} atlas building model?} In Sec.~\ref{sec:method}, we hypothesized that a \emph{forward} atlas building model is more robust than a \emph{backward} model because a \emph{forward} model evaluates the atlas-to-image similarity difference in a fixed image space while a \emph{backward} model evaluates in the evolving atlas space. Rows E and F in Tab.~\ref{tab:atlas_metrics} show that under the same conditions, a \emph{forward} model works slightly better than a \emph{backward} model; though the difference is modest. Hence, \emph{a forward model is generally preferred to a backward model when designing an atlas-building and registration framework.} 

\noindent\textbf{Q4: Can other similarity measures be used in \emph{Aladdin} to replace MSE?} We explored MSE, because of its closed-form atlas solution. However, NCC, Local NCC and mutual information could also be used. Our results for NCC\footnote{Using NCC / Local NCC will not result in a unique solution as scaling the atlas does not change the value of these measures. Hence, additional constraints might be necessary, e.g., on the average intensity of the atlas.} (row K) suggest that \emph{more advanced similarity measures could indeed result in better accuracy than MSE.}

\noindent\textbf{Q5: Should we learn an atlas or specify it as in Eq.~\eqref{eq:forward_opt_solution}?} For MSE as similarity measure, we have the choice to learn the atlas image or to specify it via Eq.~\eqref{eq:forward_opt_solution}. Rows J and K of Tab.~\ref{tab:atlas_metrics} show that using Eq.~\eqref{eq:forward_opt_solution} to specify the atlas is slightly better than learning it. This might be because learning the atlas image \emph{and} the network weights is harder than only learning the network weights. But the difference is small. If the similarity measure is not MSE, we can not use Eq.~\eqref{eq:forward_opt_solution}. In fact, there may be no closed-form solution. As indicated in \textbf{Q4}, using a more advance similarity measure, e.g. NCC, can increase registration accuracy (see row K of Tab.~\ref{tab:atlas_metrics}).
\emph{Hence, learning an atlas can be beneficial when using an advanced similarity measure, which can improve accuracy, but may not allow for a closed-form solution.}

\vspace{-2mm}
\section{Conclusion, limitations, and future work}
\label{sec:conclusion}
We introduced a joint atlas building and diffeomorphic registration learning framework (\emph{Aladdin}) that uses pairwise image alignment losses to improve registration accuracy. \emph{Aladdin} is based on an SVF transformation model and results in diffeomorphic transformation maps (between the atlas and the images) which can naturally capture affine and nonparametric transformation components, thereby avoiding the need for affine pre-registrations. We also studied atlas evaluation measures and proposed a reliable atlas evaluation measure (i.e., atlas-as-a-bridge). Our results on knee MR images show that our method achieves better performance than popular atlas building approaches. 
A limitation of \emph{Aladdin} is that it does not explicitly impose regularization on atlas scale, position, and rotation due to the second-order regularization term which does not penalize affine transformations. Empirically, the fuzzy initialization of the atlas (i.e. averaged across all training images) mostly determines the final atlas position. It would be interesting to explore the influence of the initial atlas position on the obtained atlas. Note that an atlas image initialization is usually necessary in atlas building methods. 

{\scriptsize
\noindent
\textbf{Acknowledgements.} This work was supported by NIH 1R01AR072013; it expresses the views of the authors, not of NIH. Data and research tools used in this manuscript were obtained / analyzed from the controlled access datasets distributed from the Osteoarthritis Initiative (OAI), a data repository housed within the NIMH Data Archive. OAI is a collaborative informatics system created by NIMH and NIAMS to provide a worldwide resource for biomarker identification, scientific investigation and OA drug development. Dataset identifier: NIMH Data Archive Collection ID: 2343.
}
\newpage

{\small
\bibliographystyle{ieee_fullname}
\bibliography{aaai22.bib}
}

\newpage
\appendix

\onecolumn

{ 
\section*{\centering Aladdin: Joint Atlas Building and Diffeomorphic Registration Learning with Pairwise Alignment \\Supplementary Material}
}
\noindent
This supplementary material provides proofs and additional details for our approach. Specifically, \supp~\ref{sec:variational_calculus} derives the G\^{a}teaux variations of our atlas building models leading to the closed-form solution of Eqs.~(\ref{eq:backward_opt_solution}-\ref{eq:forward_opt_solution}); \supp~\ref{sec:additonal_related_work} discusses the optimization-based atlas building and registration literature in detail; \supp~\ref{sec:proof_of_svf} presents the proof of the SVF-based Euler-Lagrange equations; \supp~\ref{sec:details_and_hyperparameters} includes details about experimental settings and hyperparameter tuning.

\section{G\^{a}teaux variation of \emph{backward} and \emph{forward} atlas building models}
\label{sec:variational_calculus}

As discussed in Sec.~\ref{sec:related_work}, we can deduct the closed-form solutions for the atlas in both \emph{backward} and \emph{forward} models via optimization theory. For simplicity, we define $\mathcal{L}_{sim}$ as the squared $L_2$ norm, i.e., $\mathcal{L}_{sim}(I, J) = \|I-J\|^2_{2} = \langle I-J, I-J \rangle = \int_{\Omega} (I(x)-J(x))^2~dx$, where $\Omega$ is the domain, $x\in \Omega$ is the position, and $\langle\cdot,\cdot\rangle$ is the usual $L_2$-product for square integrable vector-fields on $\Omega$. Denote the energy functional of Eq.~\eqref{eq:backward_model} as $E_1$ and Eq.~\eqref{eq:forward_model} as $E_2$, i.e.,
\begin{align}
    E_1(\mathcal{I}) &= \sum_{i=1}^N \int_{\Omega}(\mathcal{I}(x) - I_i \circ \Phi^{-1}_{\alpha_i}(x))^2~dx + \lambda \mathcal{L}_{reg}(\Phi^{-1}_{\alpha_i})\,, \\
    E_2(\mathcal{I}) &= \sum_{i=1}^N \int_{\Omega}(\mathcal{I} \circ \Phi^{-1}_{\beta_i}(x) - I_i(x))^2~dx + \lambda \mathcal{L}_{reg}(\Phi^{-1}_{\beta_{i}})\,.
\end{align}
By G\^{a}teaux variation w.r.t. $\mathcal{I}$, we have
\begin{align}
    \delta E_1(\mathcal{I}; d\mathcal{I}) &= \frac{\partial}{\partial \epsilon} E_1 (\mathcal{I} + \epsilon d\mathcal{I})|_{\epsilon=0}\\
    &= \frac{\partial}{\partial \epsilon} \sum_{i=1}^N \Big( \int_{\Omega} ( \mathcal{I}(x)+\epsilon d\mathcal{I}(x) - I_i \circ \Phi^{-1}_{\alpha_i}(x) )^2~dx+ \mathcal{L}_{reg}(\Phi^{-1}_{\alpha_i}) \Big)|_{\epsilon=0}\\
    &= \sum_{i=1}^N \int_{\Omega} 2\Big( \mathcal{I}(x) - I_i \circ \Phi^{-1}_{\alpha_i}(x)\Big) d\mathcal{I}(x)~dx\\
    &= 2 \langle \sum_{i=1}^N \mathcal{I} - I_i \circ \Phi^{-1}_{\alpha_i}, d\mathcal{I}  \rangle \overset{!}{=} 0\,,\, \forall d\mathcal{I}\,,\\
    \delta E_2(\mathcal{I}; d\mathcal{I}) &= \frac{\partial}{\partial \epsilon} E_2 (\mathcal{I} + \epsilon d\mathcal{I})|_{\epsilon=0}\\
    &= \frac{\partial}{\partial \epsilon} \sum_{i=1}^N \Big( \int_{\Omega} ( (\mathcal{I}+\epsilon d\mathcal{I}) \circ \Phi^{-1}_{\beta_i}(x) - I_i(x)  )^2~dx+ \mathcal{L}_{reg}(\Phi^{-1}_{\beta_i}) \Big)|_{\epsilon=0}\\
    &= \sum_{i=1}^N \int_{\Omega} 2\Big( (\mathcal{I} \circ \Phi^{-1}_{\beta_i}(x) - I_i(x))\Big) d\mathcal{I} \circ \Phi^{-1}_{\beta_i}(x)~dx\\
    &= 2\sum_{i=1}^N \int_{\Omega} \Big(\mathcal{I} \circ \Phi_{\beta_i}^{-1}(x) - I_i \circ \Phi_{\beta_i} \circ \Phi_{\beta_i}^{-1}(x)\Big) d\mathcal{I} \circ \Phi_{\beta_i}^{-1}(x)~dx\\
    &\overset{(a)}{=} 2\sum_{i=1}^N \int_{\Omega^{'}} \Big(\mathcal{I}(y) - I_i \circ \Phi_{\beta_i}(y) \Big) d\mathcal{I}(y)|D\Phi_{\beta_i}(y)|~dy\\
    &\overset{(b)}{=} 2\sum_{i=1}^N \int_{\Omega} \Big(\mathcal{I}(y) - I_i \circ \Phi_{\beta_i}(y) \Big) |D\Phi_{\beta_i}(y)|d\mathcal{I}(y)~dy\\
    &= 2 \langle \sum_{i=1}^N (\mathcal{I} - I_i \circ \Phi_{\beta_i})|D \Phi_{\beta_i}|, d\mathcal{I}  \rangle \overset{!}{=} 0\,,\, \forall d\mathcal{I}\,,  
\end{align}
where $(a)$ corresponds to a change of variables: i.e., setting $\Phi_{\beta_i}^{-1}(x)=y$ or equivalently $\Phi_{\beta_i}(y)=x$, which results in the Jacobian change of variables $|D\Phi_{\beta_i}(y)|dy = dx$. For $(b)$, the transformation $\Phi^{-1}_{\beta}$: $\Omega \rightarrow \Omega$, where $\Omega \subseteq \mathbb{R}^d$ ($d=2$ for 2D or $d=3$ for 3D). Hence, changing variables does not change the domain, i.e. $\Omega^{'} = \Omega$. 

\section{Related work: optimization-based atlas building and registration approaches}
\label{sec:additonal_related_work}
This section briefly introduces optimization-based atlas building and registration approaches.

\textbf{Optimization-based \emph{backward} atlas building and registration models:} Bhatia~\etal~\cite{bhatia2004consistent} build an atlas by forcing the sum of deformations from all images to be zero and showed good performance in a small deformation setting. To address image datasets with large deformations, Lorenzen~\etal~\cite{lorenzen2005unbiased} and Avants~\etal~\cite{avants2004geodesic} use an unbiased atlas construction scheme in the space of diffeomorphisms via the large deformation diffeomorphic metric mapping (LDDMM) model~\cite{beg2003variational}. As an extension, Lorenzen~\etal~\cite{lorenzen2006multi} extended their work to multi-modal image set registration and multi-class atlas formation by minimizing the Kullback-Leibler divergence between the estimated posteriors in a Bayesian framework. Bhatia~\etal~\cite{bhatia2007groupwise} introduced an iterative Expectation-Maximization (EM) framework to simultaneously improve both the alignment of images to their average image, as well as the segmentation of structures in the average space. Van~\etal~\cite{van2008encoding} use mesh-based representations to generalize a probabilistic atlas building approach to a joint registration and atlas estimation Bayesian inference model, which automatically determines the optimal amount of spatial blurring, the best deformation field flexibility, and the most compact mesh representation. Fletcher~\etal~\cite{fletcher2009geometric} deveoped a robust brain atlas estimation technique based on the geometric median in the LDDMM framework. To avoid building a fuzzy atlas, Wu~\etal~\cite{wu2010groupwise-a,wu2010groupwise-b} proposed to average the aligned images according to anatomical shape and distances of local patches, instead of direcly using an intensity average.
Wang~\etal~\cite{wang2010groupwise} decompose a large-scale groupwise registration problem into a series of small-scale problems, which are easier to solve and thus help registration robustness.
Jia~\etal~\cite{jia2010absorb} proposed a hierarchical groupwise registration framework termed ABSORB, which bundles similar images thus reducing registration errors and generating smooth registration paths. 
Debroux~\etal~\cite{debroux2020variational} proposed a variational model for joint segmentation, registration and atlas generation.

\textbf{Optimization-based \emph{forward} atlas building and registration models:} The \emph{forward} atlas building model is motivated by the notion of deformable templates introduced in~\cite{grenander1993general}. Given a template image $I$, an entire family of new images with similar anatomical structures is modeled as the orbit of a group of diffeomorphisms $\mathcal{G}$, i.e., $Orb(I) = \{I \circ \phi^{-1}: \phi \in \mathcal{G} \}$. 
Ma~\etal~\cite{ma2008bayesian,ma2010bayesian} formulated atlas building in a Bayesian framework and use Mode Approximation of the EM algorithm to build the atlas.  Durrleman~\etal~\cite{durrleman2008forward} apply a forward model to build atlases for curves and surfaces.  Zhang~\etal~\cite{zhang2013bayesian} proposed a generative model to jointly estimate the registration regularity, noise variance, and the atlas. Singh~\etal~\cite{singh2013vector} use a vector initial momentum parameterization of diffeomorphisms for atlas construction.  To accommodate complex structural differences across a heterogeneous group of images, Zhang~\etal~\cite{zhang2015mixture} proposed a generative Gaussian mixture model for diffeomorphic multi-atlas building.

\section{SVF based Euler-Lagrange equation}
\label{sec:proof_of_svf}
This section discusses the derivation of the Euler-Lagrange equations of the regularized stationary velocity field \emph{forward} atlas building model (i.e. Eq.~\eqref{eq:forward_svf_atlas_building_with_pairwise_alignment}).

\begin{lemma}
\label{lemma:derivative_of_deformation}
The variation of transformation $\Phi_{s,t}^v$ when $v$ is perturbed along $h$ is given by:
\begin{equation}
\begin{split}
    \partial_h \Phi_{s,t}^v &= \lim_{\epsilon \rightarrow 0} \frac{\Phi_{s,t}^{v+\epsilon h} - \Phi_{s,t}^{v}}{\epsilon}\\
    &= D\Phi_{s,t}^v \int_s^t (D\phi_{s,u}^{v})^{-1}h \circ \Phi_{s,u}^{v}~du \,.
\end{split}
\end{equation}
\begin{proof}
See~\cite{beg2003variational}.
\end{proof}

\end{lemma}

\begin{theorem}
\label{thm:ele_atlas}
Given a continuous differentiable idealized atlas image $\mathcal{I}$ and a population of noisy observed anatomies $\{I_i\}$ ($i = 1, ..., N$), the $\{v_i^*\}$ minimizing the following energy functional
\begin{equation}
    E_0(\{v_i\}) = \sum_{i=1}^N \big[ \lambda \text{Reg}(v_i) + \| \mathcal{I} \circ \Phi_{1, 0}^{v_i} - I_i \|^2_{2} \big]
\end{equation}
satisfy the Euler-Lagrange equation
\begin{equation}
\label{eq:ele_atlas}
    \lambda \nabla_{v_i^{*}} \text{Reg}(v_i^{*}) - 2 \int_0^1 \Big( |D\Phi_{t,1}^{v_i^*}| (\mathcal{I} \circ \Phi^{v_i^*}_{t, 0} - I_i \circ \Phi^{v_i^*}_{t, 1}) \nabla(\mathcal{I} \circ \Phi^{v_i^*}_{t, 0}) \Big)~dt = 0, \forall i\,.
\end{equation}
\end{theorem}

\begin{proof}
The Euler-Lagrange equation associated to energy functional $E_0(\{v_i\})$ is obtained by setting the Fr\'{e}chet derivative $\nabla_{v_i} E_0(\{v_i\})$ to zero for each $i$. Let the velocity $v_i$ be perturbed by an $\epsilon$ amount along direction $h_i$. The Fr\'{e}chet derivative $\nabla_{v_i} E_0(\{v_i\})$ can be computed from the G\^{a}teaux variation $\partial_{h_i} E_0(\{v_i\})$ by
\begin{equation}
\begin{split}
    \partial_{h_i} E_0(\{v_i\}) &= \lim_{\epsilon \rightarrow 0} \frac{E_0(\{v_i+\epsilon h_i\}) - E_0(\{v_i\})}{\epsilon}\\
    &= \langle \nabla_{v_i} E_0, h_i \rangle\,.
\end{split}
\end{equation}
Denote $E_{reg}(v_i) = \text{Reg}(v_i)$ and $E_{sim}(v_i) = \| \mathcal{I} \circ \Phi_{1, 0}^{v_i} - I_i \|^2_{2}$, then we have $\partial_{h_i} E_0(\{v_i\}) = \lambda\partial_{h_i} E_{reg}(v_i) + \partial_{h_i} E_{sim}(v_i)$.

The variation of $E_{reg}(v_i)$ is
\begin{equation*}
    \partial_{h_i} E_{reg}(v_i) = \langle \nabla_{v_i}\text{Reg}(v_i), h_i \rangle\,.
\end{equation*}

Similar to \cite{beg2003variational}, the variation of $E_{sim}(v_i)$ is
\begin{align*}
    \partial_{h_i} E_{sim}(v_i) &= 2\big\langle \mathcal{I} \circ \Phi_{1,0}^{v_i} - I_i, D\mathcal{I}\circ\Phi_{1,0}^{v_i}\partial_{h_i}\Phi_{1,0}^{v_i} \big\rangle\\
    &\overset{(a)}{=} 2\Big\langle \mathcal{I} \circ \Phi_{1,0}^{v_i} - I_i, D\mathcal{I} \circ \Phi_{1,0}^{v_i} \times \Big ( -D\Phi_{1,0}^{v_i} \int_0^1 (D\Phi_{1,t}^{v_i})^{-1} h_i \circ \Phi_{1,0}^{v_i}~dt \Big) \Big\rangle\\
    &\overset{(b)}{=} -2\int_0^1 \big\langle \mathcal{I} \circ \Phi_{1,0}^{v_i} - I_i, D(\mathcal{I}\circ\Phi_{1,0}^{v_i}) \times (D\Phi_{1,t}^{v_i})^{-1}h_i\circ\Phi^{v_i}_{1,t} \big\rangle~dt\\
    &= -2\int_0^1 \big\langle (\mathcal{I} \circ \Phi_{t,0}^{v_i} - I_i\circ \Phi_{t,1}^{v_i})\circ\Phi_{1,t}^{v_i}, D(\mathcal{I}\circ\Phi_{t,0}^{v_i} \circ \Phi_{1,t}^{v_i}) \times (D\Phi_{1,t}^{v_i})^{-1}h_i\circ\Phi^{v_i}_{1,t} \big\rangle~dt\\
    &\overset{(c)}{=} -2\int_0^1 \big\langle |D\Phi_{t,1}^{v_i}| (\mathcal{I} \circ \Phi_{t,0}^{v_i} - I_i \circ \Phi_{t,1}^{v_i}), D(\mathcal{I} \circ \Phi_{t,0}^{v_i})h_i \big\rangle~dt\\
    &\overset{(d)}{=} -2\int_0^1 \big\langle |D\Phi_{t,1}^{v_i}| (\mathcal{I} \circ \Phi_{t,0}^{v_i} - I_i \circ \Phi_{t,1}^{v_i}) \nabla(\mathcal{I} \circ \Phi_{t,0}^{v_i}), h_i \big\rangle~dt\\
    &= \Big\langle -2\int_0^1 \Big( |D\Phi_{t,1}^{v_i}| (\mathcal{I} \circ \Phi_{t,0}^{v_i} - I_i \circ \Phi_{t,1}^{v_i}) \nabla(\mathcal{I} \circ \Phi_{t,0}^{v_i}) \Big)~dt, h_i \Big\rangle
\end{align*}
where $(a)$ refers to substituting $\partial_{h_i}\Phi_{1,0}^{v_i}$ based on Lemma~\ref{lemma:derivative_of_deformation}; $(b)$ refers to rewriting $D(\mathcal{I} \circ \Phi_{1,0}^{v_i}) = D\mathcal{I}\circ\Phi_{1,0}^{v_i}D\Phi_{1,0}^{v_i}$; $(c)$ is the chain rule ($D(\mathcal{I}\circ\Phi_{t,0}^{v_i} \circ \Phi_{1,t}^{v_i}) = D_{\Phi_{1,t}^{v_i}}(\mathcal{I}\circ\Phi_{t,0}^{v_i}) D\Phi_{1,t}^{v_i}$) and the Jacobian change of variables. Denote $\Phi_{1,t}^{v_i}(x) = y$, then $\Phi_{t,1}^{v_i}(y) = x$ and the Jacobian change of variables is $|D\Phi_{t,1}^{v_i}(y)|dy=dx$; $(d)$ follows from writing the transpose and changing the notation $\nabla (\mathcal{I} \circ \Phi_{t,0}^{v_i}) = D(\mathcal{I} \circ \Phi_{t,0}^{v_i})^{T}$.

Collecting terms, the Fr\'{e}chet derivative $\nabla_{v_i} E_0(\{v_i\})$ is
\begin{equation}
    \nabla_{v_i} E_0(\{v_i\}) = \lambda \nabla_{v_i} \text{Reg}(v_i) - 2 \int_0^1 \Big( |D\Phi_{t,1}^{v_i}|  (\mathcal{I} \circ \Phi^{v_i}_{t, 0} - I_i \circ \Phi^{v_i}_{t, 1}) \nabla(\mathcal{I} \circ \Phi^{v_i}_{t, 0}) \Big)~dt\,.
\end{equation}
This equation yields the Euler-Lagrange Eq.~\eqref{eq:ele_atlas}.
\end{proof}

\begin{theorem}
\label{thm:ele_atlas_atlas_image_together}
Given a continuous differentiable idealized atlas image $\mathcal{I}$ and a population of noisy observed anatomies $\{I_i\}$ ($i = 1, ..., N$), the $\{v_i^*\}$ minimizing the following energy functional
\begin{equation}
\begin{split}
    E_1(\{v_i\}) = &\sum_{(i, j) \in \Gamma} \Big[ \lambda \text{Reg}(v_i) + \| \mathcal{I} \circ \Phi_{1, 0}^{v_i} - I_i \|^2_{2} + \lambda \text{Reg}(v_j) + \| \mathcal{I} \circ \Phi_{1, 0}^{v_j} - I_j \|^2_{2} \\
    & + \gamma_1 \| I_i \circ \Phi_{0,1}^{v_i} - I_j \circ \Phi_{0,1}^{v_j} \|^2_{2} \\
    & + \gamma_2 \Big( \| I_i \circ \Phi_{0,1}^{v_i} \circ \Phi_{1,0}^{v_j} - I_j \|^2_{2} + \| I_j \circ \Phi_{0,1}^{v_j} \circ \Phi_{1,0}^{v_i} - I_i \|^2_{2} \Big) \Big]
\end{split}
\end{equation}
satisfy the Euler-Lagrange equation
\begin{equation}
\begin{split}
\label{eq:ele_pairwise_atlas_image_together_app}
    (N-1) \lambda \nabla_{v_i^*} \text{Reg}(v_i^*) &-  2(N-1) \int_0^1 \Big( |D\Phi_{t,1}^{v_i^*}| (\mathcal{I} \circ \Phi^{v_i^*}_{t, 0} - I_i \circ \Phi^{v_i^*}_{t, 1}) \nabla(\mathcal{I} \circ \Phi^{v_i^*}_{t, 0}) \Big)~dt \\
    &- 2N\gamma_1 \int_0^1 \big( |D\Phi_{t,0}^{v_i^*}|(\frac{\sum_{j=1}^N I_i \circ \Phi_{0,1}^{v_j^*}}{N} \circ \Phi_{t,0}^{v_i^*} - I_i \circ \Phi_{t,1}^{v_i^*}) \nabla (I_i \circ \Phi_{t,1}^{v_i^*}) \big)~dt \\
    &- 2\gamma_2 \int_0^1 \Big( \big( \sum_{j=1}^N |D\Phi_{0,1}^{v^*_j}| \big)|D\Phi_{t,0}^{v^*_i}|( \frac{\sum_{j=1}^N |D\Phi_{0,1}^{v^*_j}| I_j \circ \Phi_{0,1}^{v^*_j}}{\sum_{j=1}^N |D\Phi_{0,1}^{v^*_j}|} \circ \Phi_{t,0}^{v^*_i} - I_i \circ \Phi_{t,1}^{v^*_i}) \nabla (I_i \circ \Phi_{t,1}^{v^*_i}) \Big)~dt\\
    &- 2\gamma_2 \int_0^1 \sum_{j=1}^N |D\Phi_{t,1}^{v^*_i}| (I_j \circ \Phi_{0,1}^{v^*_j} \circ \Phi_{t,0}^{v^*_i} - I_i \circ \Phi_{t,1}^{v^*_i}) \nabla(I_j \circ \Phi_{0,1}^{v^*_j} \circ \Phi_{t,0}^{v^*_i})~dt = 0\,, \forall i\,.
\end{split}
\end{equation}
\end{theorem}

\begin{proof}
Similar to the proof of Theorem~\ref{thm:ele_atlas}, we need to calculate the Fr\'{e}chet derivative $\nabla_{v_i} E_1(\{v_i\})$. Denote $E^{atlas}_{pair}(v_i, v_j) = \sum_{(i, j) \in \Gamma} \| I_i \circ \Phi_{0,1}^{v_i} - I_j \circ \Phi_{0,1}^{v_j} \|^2_{2}$,  $E^{image}_{pair}(v_i, v_j) = \sum_{(i, j) \in \Gamma} \| I_i \circ \Phi_{0,1}^{v_i} \circ \Phi_{1,0}^{v_j} - I_j \|^2_{2}$ and $E^{image}_{pair}(v_j, v_i) = \sum_{(i, j) \in \Gamma} \| I_j \circ \Phi_{0,1}^{v_j} \circ \Phi_{1,0}^{v_i} - I_i \|^2_{2}$, then we have $\partial_{h_i} E_1(\{v_i\}) = (N-1) \lambda \partial_{h_i} E_{reg}(v_i) + (N-1) \partial_{h_i} E_{sim}(v_i) + \gamma_1 \partial_{h_i} E^{atlas}_{pair}(v_i, v_j) + \gamma_2 \partial_{h_i} E^{image}_{pair}(v_i, v_j) + \gamma_2 \partial_{h_i} E^{image}_{pair}(v_j, v_i)$. 

For conciseness, we ignore the calculation of $\partial_{h_i} E_{reg}(v_i)$ and $\partial_{h_i} E_{sim}(v_i)$, as they have been obtained in the proof of Theorem~\ref{thm:ele_atlas}.

The variation of $E^{atlas}_{pair}(v_i, v_j)$ is
\begin{align*}
    \partial_{h_i} E^{atlas}_{pair}(v_i, v_j) &= \sum_{(i, j) \in \Gamma} 2\big\langle I_i \circ \Phi_{0,1}^{v_i} - I_j \circ \Phi_{0,1}^{v_j}, DI_i \circ \Phi_{0,1}^{v_i}\partial_{h_i}\Phi_{0,1}^{v_i} \big\rangle \\
    &= 2(N-1) \big\langle I_i \circ \Phi_{0,1}^{v_i} - \frac{\sum_{j=1, j \neq i}^N I_j \circ \Phi_{0,1}^{v_j}}{N-1}, DI_i \circ \Phi_{0,1}^{v_i}\partial_{h_i}\Phi_{0,1}^{v_i} \big\rangle\\
    &= 2N \big\langle I_i \circ \Phi_{0,1}^{v_i} - \frac{\sum_{j=1}^N I_j \circ \Phi_{0,1}^{v_j}}{N}, DI_i \circ \Phi_{0,1}^{v_i}\partial_{h_i}\Phi_{0,1}^{v_i} \big\rangle\\
    &\overset{(a)}{=}2N \big\langle I_i \circ \Phi_{0,1}^{v_i} - \hat{I}, DI_i \circ \Phi_{0,1}^{v_i}\partial_{h_i}\Phi_{0,1}^{v_i} \big\rangle\\
    &\overset{(b)}{=} 2N \Big\langle I_i \circ \Phi_{0,1}^{v_i} - \hat{I}, DI_i \circ \Phi_{0,1}^{v_i} \times \Big( D\Phi_{0,1}^{v_i} \int_0^1 (D\Phi_{0,t}^{v_i})^{-1} h_i \circ \Phi_{0,t}^{v_i}~dt \Big)  \Big\rangle\\
    &\overset{(c)}{=} 2N \int_0^1 \big\langle I_i \circ \Phi_{0,1}^{v_i} - \hat{I}, D(I_i \circ \Phi_{0,1}^{v_i}) \times (D\Phi_{0,t}^{v_i})^{-1} h_i \circ \Phi_{0,t}^{v_i} \big\rangle~dt\\
    &= 2N \int_0^1 \big\langle (I_i \circ \Phi_{t,1}^{v_i} - \hat{I} \circ \Phi_{t,0}^{v_i}) \circ \Phi_{0,t}^{v_i}, D(I_i \circ \Phi_{t,1}^{v_i} \circ \Phi_{0,t}^{v_i}) \times (D\Phi_{0,t}^{v_i})^{-1} h_i \circ \Phi_{0,t}^{v_i} \big\rangle~dt\\
    &\overset{(d)}{=} 2N \int_0^1 \big\langle |D\Phi_{t,0}^{v_i}|(I_i \circ \Phi_{t,1}^{v_i} - \hat{I} \circ \Phi_{t,0}^{v_i}), D(I_i \circ \Phi_{t,1}^{v_i})h_i \big\rangle~dt\\
    &\overset{(e)}{=} 2N \int_0^1 \big\langle |D\Phi_{t,0}^{v_i}|(I_i \circ \Phi_{t,1}^{v_i} - \hat{I} \circ \Phi_{t,0}^{v_i}) \nabla (I_i \circ \Phi_{t,1}^{v_i}), h_i \big\rangle~dt\\
    &= \Big\langle -2N \int_0^1 \big( |D\Phi_{t,0}^{v_i}|( \hat{I} \circ \Phi_{t,0}^{v_i} - I_i \circ \Phi_{t,1}^{v_i} ) \nabla (I_i \circ \Phi_{t,1}^{v_i}) \big)~dt, h_i \Big\rangle
\end{align*}
where $(a)$ is using $\hat{I}$ to represent $\frac{\sum_{j=1}^N I_j \circ \Phi_{0,1}^{v_j}}{N}$; $(b)$ corresponds to substituting $\partial_{h_i}\Phi_{0,1}^{v_i}$ based on Lemma~\ref{lemma:derivative_of_deformation}; $(c)$ amounts to rewriting $D(I_i \circ \Phi_{0,1}^{v_i}) = DI_i\circ\Phi_{0,1}^{v_i}D\Phi_{0,1}^{v_i}$; $(d)$ is the chain rule ($D(I_i \circ \Phi_{t,1}^{v_i} \circ \Phi_{0,t}^{v_i}) = D_{\Phi_{0,t}^{v_i}}(I_i \circ \Phi_{t,1}^{v_i})D \Phi_{0,t}^{v_i}$) the Jacobian change of variables. Denote $\Phi_{0,t}^{v_i}(x) = y$, then $\Phi_{t,0}^{v_i}(y) = x$ and the Jacobian change of variables is $|D\Phi_{t,0}^{v_i}(y)|dy=dx$; $(e)$ results from writing the transpose and changing the notation $\nabla (I_i \circ \Phi_{t,1}^{v_i}) = D(I_i \circ \Phi_{t,1}^{v_i})^{T}$.

Before calculating the variation of $E^{image}_{pair}(v_i, v_j)$, we first reformat the equation
\begin{align*}
    E^{image}_{pair}(v_i, v_j) &= \| I_i \circ \Phi_{0,1}^{v_i} \circ \Phi_{1,0}^{v_j} - I_j \|^2_{2}\\
    &= \big\langle I_i \circ \Phi_{0,1}^{v_i} \circ \Phi_{1,0}^{v_j} - I_j, I_i \circ \Phi_{0,1}^{v_i} \circ \Phi_{1,0}^{v_j} - I_j \big\rangle\\
    &= \big\langle (I_i \circ \Phi_{0,1}^{v_i} -  I_j \circ \Phi_{0,1}^{v_j}) \circ \Phi_{1,0}^{v_j}, (I_i \circ \Phi_{0,1}^{v_i} -  I_j \circ \Phi_{0,1}^{v_j}) \circ \Phi_{1,0}^{v_j} \big\rangle\\
    &\overset{(a)}{=} \big\langle |D\Phi_{0,1}^{v_j}|(I_i \circ \Phi_{0,1}^{v_i} -  I_j \circ \Phi_{0,1}^{v_j}), (I_i \circ \Phi_{0,1}^{v_i} -  I_j \circ \Phi_{0,1}^{v_j}) \big\rangle 
\end{align*}
where $(a)$ is the Jacobian change of variables. Denote $\Phi_{1,0}^{v_j}(x) = y$, then $\Phi_{0,1}^{v_j}(y) = x$ and the Jacobian change of variables is $|D\Phi_{0,1}^{v_j}(y)|dy=dx$.
Hence the variation of $E^{image}_{pair}(v_i, v_j)$ is
\begin{align*}
    \partial_{h_i} E^{image}_{pair}(v_i, v_j) &= 2\sum_{(i, j) \in \Gamma} \big\langle |D\Phi_{0,1}^{v_j}|(I_i \circ \Phi_{0,1}^{v_i} -  I_j \circ \Phi_{0,1}^{v_j}), DI_i \circ \Phi_{0,1}^{v_i}\partial_{h_i}\Phi_{0,1}^{v_i} \big\rangle\\
    &= 2\Big\langle \big( \sum_{j=1, j \neq i}^N |D\Phi_{0,1}^{v_j}| \big) (I_i \circ \Phi_{0,1}^{v_i} - \frac{\sum_{j=1, j \neq i}^N |D\Phi_{0,1}^{v_j}| I_j \circ \Phi_{0,1}^{v_j}}{\sum_{j=1, j \neq i}^N |D\Phi_{0,1}^{v_j}|}), DI_i \circ \Phi_{0,1}^{v_i}\partial_{h_i}\Phi_{0,1}^{v_i} \Big\rangle\\
    &= 2\Big\langle \big( \sum_{j=1}^N |D\Phi_{0,1}^{v_j}| \big) (I_i \circ \Phi_{0,1}^{v_i} - \frac{\sum_{j=1}^N |D\Phi_{0,1}^{v_j}| I_j \circ \Phi_{0,1}^{v_j}}{\sum_{j=1}^N |D\Phi_{0,1}^{v_j}|}), DI_i \circ \Phi_{0,1}^{v_i}\partial_{h_i}\Phi_{0,1}^{v_i} \Big\rangle\\
    &\overset{(a)}{=} 2\Big\langle \big( \sum_{j=1}^N |D\Phi_{0,1}^{v_j}| \big) (I_i \circ \Phi_{0,1}^{v_i} - \tilde{I}), DI_i \circ \Phi_{0,1}^{v_i}\partial_{h_i}\Phi_{0,1}^{v_i} \Big\rangle\\
    &\overset{(b)}{=} 2\Big\langle \big( \sum_{j=1}^N |D\Phi_{0,1}^{v_j}| \big)(I_i \circ \Phi_{0,1}^{v_i} - \tilde{I}), DI_i \circ \Phi_{0,1}^{v_i} \times \Big( D\Phi_{0,1}^{v_i} \int_0^1 (D\Phi_{0,t}^{v_i})^{-1} h_i \circ \Phi_{0,t}^{v_i}~dt \Big)  \Big\rangle\\
    &\overset{(c)}{=} 2\int_0^1 \big\langle \big( \sum_{j=1}^N |D\Phi_{0,1}^{v_j}| \big)(I_i \circ \Phi_{0,1}^{v_i} - \tilde{I}), D(I_i \circ \Phi_{0,1}^{v_i}) \times (D\Phi_{0,t}^{v_i})^{-1} h_i \circ \Phi_{0,t}^{v_i} \big\rangle~dt\\
    &= 2\int_0^1 \big\langle \big( \sum_{j=1}^N |D\Phi_{0,1}^{v_j}| \big) (I_i \circ \Phi_{t,1}^{v_i} - \tilde{I} \circ \Phi_{t,0}^{v_i}) \circ \Phi_{0,t}^{v_i}, D(I_i \circ \Phi_{t,1}^{v_i} \circ \Phi_{0,t}^{v_i}) \times (D\Phi_{0,t}^{v_i})^{-1} h_i \circ \Phi_{0,t}^{v_i} \big\rangle~dt\\
    &\overset{(d)}{=} 2\int_0^1 \big\langle \big( \sum_{j=1}^N |D\Phi_{0,1}^{v_j}| \big)|D\Phi_{t,0}^{v_i}|(I_i \circ \Phi_{t,1}^{v_i} - \tilde{I} \circ \Phi_{t,0}^{v_i}), D(I_i \circ \Phi_{t,1}^{v_i})h_i \big\rangle~dt\\
    &\overset{(e)}{=} 2\int_0^1 \big\langle \big( \sum_j^N |D\Phi_{0,1}^{v_j}| \big)|D\Phi_{t,0}^{v_i}|(I_i \circ \Phi_{t,1}^{v_i} - \tilde{I} \circ \Phi_{t,0}^{v_i}) \nabla (I_i \circ \Phi_{t,1}^{v_i}), h_i \big\rangle~dt\\
    &= \Big\langle - 2\int_0^1 \Big( \big( \sum_{j=1}^N |D\Phi_{0,1}^{v_j}| \big)|D\Phi_{t,0}^{v_i}|( \tilde{I} \circ \Phi_{t,0}^{v_i} - I_i \circ \Phi_{t,1}^{v_i} ) \nabla (I_i \circ \Phi_{t,1}^{v_i}) \Big)~dt, h_i \Big\rangle
\end{align*}
where $(a)$ is using $\tilde{I}$ to represent $\frac{\sum_{j=1}^N |D\Phi_{0,1}^{v_j}| I_j \circ \Phi_{0,1}^{v_j})}{\sum_{j=1}^N |D\Phi_{0,1}^{v_j}|}$; $(b)$ refers to substituting $\partial_{h_i}\Phi_{0,1}^{v_i}$ based on Lemma~\ref{lemma:derivative_of_deformation}; $(c)$ amounts to rewriting $D(I_i \circ \Phi_{0,1}^{v_i}) = DI_i\circ\Phi_{0,1}^{v_i}D\Phi_{0,1}^{v_i}$; $(d)$ is the chain rule ($D(I_i \circ \Phi_{t,1}^{v_i} \circ \Phi_{0,t}^{v_i}) = D_{\Phi_{0,t}^{v_i}}(I_i \circ \Phi_{t,1}^{v_i}) D \Phi_{0,t}^{v_i}$) and the Jacobian change of variables. Denote $\Phi_{0,t}^{v_i}(x) = y$, then $\Phi_{t,0}^{v_i}(y) = x$ and the Jacobian change of variables is $|D\Phi_{t,0}^{v_i}(y)|dy=dx$; $(e)$ refers to writing the transpose and changing the notation $\nabla (I_i \circ \Phi_{t,1}^{v_i}) = D(I_i \circ \Phi_{t,1}^{v_i})^{T}$.

The variation of $E^{image}_{pair}(v_j, v_i)$ is
\begin{align*}
    \partial_{h_i} E^{image}_{pair}(v_j, v_i) 
    &= \sum_{(i, j) \in \Gamma} 2\big\langle I_j \circ \Phi_{0,1}^{v_j} \circ \Phi_{1,0}^{v_i} - I_i, D(I_j \circ \Phi_{0,1}^{v_j}) \circ \Phi_{1,0}^{v_i} \partial_{h_i}\Phi_{1,0}^{v_i} \big\rangle\\
    &\overset{(a)}{=} \sum_{j=1, j \neq i}^N 2\Big\langle I_j \circ \Phi_{0,1}^{v_j} \circ \Phi_{1,0}^{v_i} - I_i, D(I_j \circ \Phi_{0,1}^{v_j}) \circ \Phi_{1,0}^{v_i} \times \Big(-D\Phi_{1,0}^{v_i}\int_0^1 (D\Phi_{1,t}^{v_i})^{-1} h_i \circ \Phi_{1,t}^{v_i}~dt \Big) \Big\rangle\\
    &= \sum_{j=1}^N 2\Big\langle I_j \circ \Phi_{0,1}^{v_j} \circ \Phi_{1,0}^{v_i} - I_i, D(I_j \circ \Phi_{0,1}^{v_j}) \circ \Phi_{1,0}^{v_i} \times \Big(-D\Phi_{1,0}^{v_i}\int_0^1 (D\Phi_{1,t}^{v_i})^{-1} h_i \circ \Phi_{1,t}^{v_i}~dt \Big) \Big\rangle\\
    &= - 2\sum_{j=1}^N \int_0^1 \big\langle I_j \circ \Phi_{0,1}^{v_j} \circ \Phi_{1,0}^{v_i} - I_i, D(I_j \circ \Phi_{0,1}^{v_j} \circ \Phi_{1,0}^{v_i}) \times (D\Phi_{1,t}^{v_i})^{-1} h_i \circ \Phi_{1,t}^{v_i} \big\rangle~dt\\
    &\overset{(b)}{=} - 2\sum_{j=1}^N \int_0^1 \big\langle (I_j \circ \Phi_{0,1}^{v_j} \circ \Phi_{t,0}^{v_i} - I_i \circ \Phi_{t,1}^{v_i}) \circ \Phi_{1,t}^{v_i}, D(I_j \circ \Phi_{0,1}^{v_j} \circ \Phi_{t,0}^{v_i} \circ \Phi_{1,t}^{v_i}) \times (D\Phi_{1,t}^{v_i})^{-1} h_i \circ \Phi_{1,t}^{v_i} \big\rangle\\
    & \overset{(c)}{=} - 2\sum_{j=1}^N \int_0^1 \big\langle |D\Phi_{t,1}^{v_i}| (I_j \circ \Phi_{0,1}^{v_j} \circ \Phi_{t,0}^{v_i} - I_i \circ \Phi_{t,1}^{v_i}), D(I_j \circ \Phi_{0,1}^{v_j} \circ \Phi_{t,0}^{v_i}) h_i \big\rangle\\
    &\overset{(d)}{=} -2 \sum_{j=1}^N \int_0^1 \big\langle |D\Phi_{t,1}^{v_i}| (I_j \circ \Phi_{0,1}^{v_j} \circ \Phi_{t,0}^{v_i} - I_i \circ \Phi_{t,1}^{v_i}) \nabla(I_j \circ \Phi_{0,1}^{v_j} \circ \Phi_{t,0}^{v_i}), h_i \big\rangle\\
    &= \Big\langle -2 \int_0^1 \sum_{j=1}^N |D\Phi_{t,1}^{v_i}| (I_j \circ \Phi_{0,1}^{v_j} \circ \Phi_{t,0}^{v_i} - I_i \circ \Phi_{t,1}^{v_i}) \nabla(I_j \circ \Phi_{0,1}^{v_j} \circ \Phi_{t,0}^{v_i})~dt, h_i \Big\rangle
\end{align*}
where $(a)$ results from substituting $\partial_{h_i}\Phi_{1,0}^{v_i}$ based on Lemma~\ref{lemma:derivative_of_deformation}; $(b)$ amounts to rewriting $D(I_j \circ \Phi_{0,1}^{v_j} \circ \Phi_{1,0}^{v_i}) = D(I_j \circ \Phi_{0,1}^{v_j}) \circ \Phi_{1,0}^{v_i}D\Phi_{1,0}^{v_i}$; $(c)$ is the chain rule ($D(I_j \circ \Phi_{0,1}^{v_j} \circ \Phi_{t,0}^{v_i} \circ \Phi_{1,t}^{v_i}) = D_{\Phi_{1,t}^{v_i}}(I_j \circ \Phi_{0,1}^{v_j} \circ \Phi_{t,0}^{v_i}) D \Phi_{1,t}^{v_i}$) and the Jacobian change of variables. Denote $\Phi_{1,t}^{v_i}(x) = y$, then $\Phi_{t,1}^{v_i}(y) = x$ and the Jacobian change of variables is $|D\Phi_{t,1}^{v_i}(y)|dy=dx$; $(d)$ is the result of writing the transpose and changing the notation $\nabla (I_j \circ \Phi_{0,1}^{v_j} \circ \Phi_{t,0}^{v_i}) = D(I_j \circ \Phi_{0,1}^{v_j} \circ \Phi_{t,0}^{v_i})^{T}$.

Collecting terms, the Fr\'{e}chet derivative $\nabla_{v_i} E_1(\{v_i\})$ is
\begin{equation}
\begin{split}
    \nabla_{v_i} E_1(\{v_i\}) = (N-1)\lambda \nabla_{v_i}\text{Reg}(v_i) &- 2(N-1) \int_0^1 \Big( |D\Phi_{t,1}^{v_i}| (\mathcal{I} \circ \Phi^{v_i}_{t, 0} - I_i \circ \Phi^{v_i}_{t, 1}) \nabla(\mathcal{I} \circ \Phi^{v_i}_{t, 0}) \Big)~dt \\
    &- 2N\gamma_1 \int_0^1 \big( |D\Phi_{t,0}^{v_i}|( \hat{I} \circ \Phi_{t,0}^{v_i} - I_i \circ \Phi_{t,1}^{v_i}) \nabla (I_i \circ \Phi_{t,1}^{v_i}) \big)~dt \\
    &- 2\gamma_2 \int_0^1 \Big( \big( \sum_{j=1}^N |D\Phi_{0,1}^{v_j}| \big)|D\Phi_{t,0}^{v_i}|(\tilde{I} \circ \Phi_{t,0}^{v_i} - I_i \circ \Phi_{t,1}^{v_i}) \nabla (I_i \circ \Phi_{t,1}^{v_i}) \Big)~dt\\
    &- 2\gamma_2 \int_0^1 \sum_{j=1}^N |D\Phi_{t,1}^{v_i}| (I_j \circ \Phi_{0,1}^{v_j} \circ \Phi_{t,0}^{v_i} - I_i \circ \Phi_{t,1}^{v_i}) \nabla(I_j \circ \Phi_{0,1}^{v_j} \circ \Phi_{t,0}^{v_i})~dt
\end{split}
\end{equation}
This equation yields the Euler-Lagrange Eq.~\eqref{eq:ele_pairwise_atlas_image_together_app}.
\end{proof}

\section{Experimental Details and Additional Results}
\label{sec:details_and_hyperparameters}
This section provides details about our experimental settings and hyperparameter tuning. To ensure fair comparisons between methods with different hyperparameter settings, we use the same random seed throughout all experiments. When training or optimizing models, we obtain the best hyperparameters for each model by tuning on the same hold-out validation dataset with grid search over a subset of hyperparameter combinations. Then the estimated best hyperparameters are used to report the performances in the test dataset. The presented results are from a single instance of each model, not averaged repeated results with different initializations and hyperparameters.

\subsection{Fig.~\ref{fig:pairwise_loss_demo} experimental details}
In this experiment, we fix the atlas and use the following parameters. The coefficient for the similarity loss is $10.0$, $\lambda=20000.0$, $\gamma_1 = 2.0$, and $\gamma_2 = 5.0$. We use \texttt{ADAM} as the optimizer to learn network parameters with a multi-step learning rate over 500 epochs. The initial learning rate is $10^{-4}$, which is multiplied by $0.1$ at the 300-th epoch and the 420-th epoch. Batch size is 2. We use MSE as the similarity loss.

\begin{figure*}[t!]
  \centering
  \includegraphics[width=0.9\linewidth]{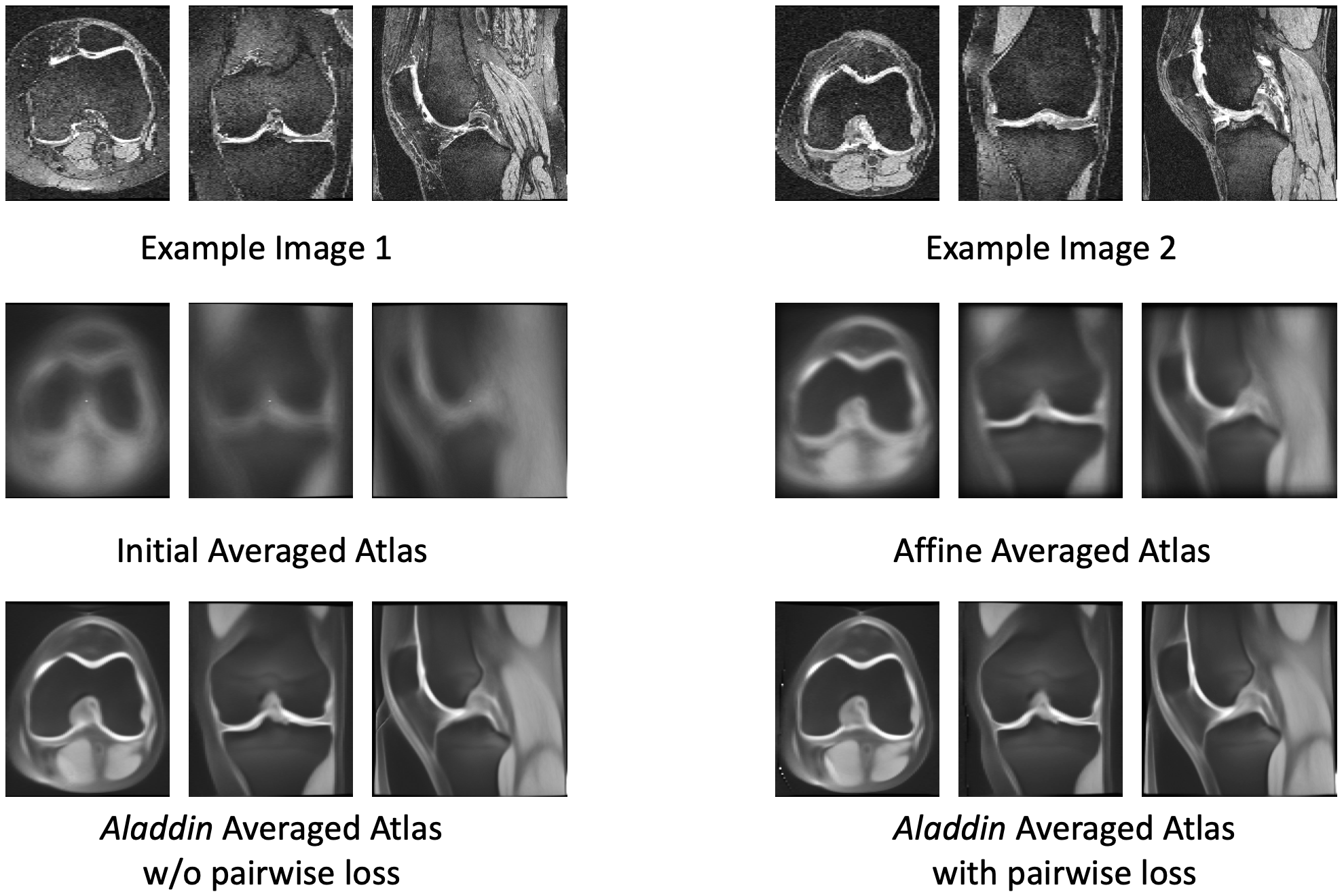}
  \caption{Examples of atlases and images in axial, coronal, and sagittal planes. \textbf{Top:} two example images. \textbf{Middle:} initial averaged atlas without and with affine pre-alignment. \textbf{Bottom:} \emph{Aladdin} results without and with pairwise alignment loss. \emph{Aladdin} obtains a relatively sharp atlas, maintaining the main structures, from the image population.}
  \label{fig:atlas_comparison}
\end{figure*}

\subsection{Tab.~\ref{tab:atlas_metrics} experimental details}
\noindent {\bf Affine Registration}: Affine registrations are performed by iteratively registering all images to the average of all images and updating the averaged image based on the resulting warped images. We use \texttt{reg\_aladin} of Nifty-Reg~\cite{rueckert1999nonrigid,modat2010fast,modat2014global,ourselin2001reconstructing} for affine registration with MSE as similarity loss. These optimization-based affine transformations also serve as the affine pre-alignments for other experiments that require affine pre-registrations.\\
{\bf Joshi~\etal~\cite{joshi2004unbiased}}: We use the \emph{Fast Symmetric Forces Demons Algorithm}~\cite{vercauteren2009diffeomorphic} (via SimpleITK) as the registration algorithm. The Gaussian smoothing standard deviation for the displacement field is set to 1.2 and the total iteration number is set to 1,000. All the other hyperparameters are default. We use MSE as the similarity loss.\\
{\bf ABSORB~\cite{jia2010absorb}}: We use \emph{Diffeomorphic Demons}~\cite{vercauteren2009diffeomorphic} (via SimpleITK) as the registration algorithm. The Gaussian smoothing standard deviation for the displacement field is set to 2.0 and the total iteration number is set to 1,000. We set the ABSORB hyperparameters as follows. We set the neighborhood size to 3, the Gaussian smoothing standard deviation for the displacement field is set to 2.0, the maximum number of levels to 3, the registration to mean to 1, histogram matching to true, and affine registration to false. All the other hyperparameters are default. We use MSE as the similarity loss.\\
{\bf Voxelmorph~\cite{balakrishnan2019voxelmorph}}: We set the regularization coefficient to 2,000 for experiments with affine pre-alignment and to 400 without affine pre-alignments. We use \texttt{ADAM} as the optimizer with  learning rate $10^{-4}$. Batch size is 4. We use MSE as the similarity loss.\\
{\bf He~\etal~\cite{he2020unsupervised}}: We implement this method based on the descriptions in \cite{he2020unsupervised}. We use a 1-step 1-iteration framework to ensure a fair comparison\footnote{If we would use a multi-step and multi-iteration framework to generate better results, to ensure fair comparisons, all the other baseline methods would also need to use a multi-step and multi-iteration approach.}. We add the atlas according to Eq.~\eqref{eq:backward_opt_solution} in the implementation because we found that without an atlas the registration performance is significantly worse than with an atlas. The coefficient for $\mathcal{L}_{grad}$ is 10, the coefficient for $\mathcal{L}_{pres}$ is 1.0, the coefficient for $\mathcal{L}_{sim}$ is 0.4, and the coefficient for $\mathcal{L}_{cycle}$ is 0.1. We use \texttt{ADAM} as the optimizer with learning rate $10^{-4}$ over 500 epochs. Batch size is 8. We use NCC as the similarity loss.\\
{\bf Dalca~\etal~\cite{dalca2019learning}}: For experiments with affine pre-alignment we set $\gamma = 0.01$, $\lambda_d = 1.0$, $\lambda_a = 100.0$, and $\sigma^2=0.5$. For experiments without affine pre-alignment we set $\gamma = 0.01$, $\lambda_d = 0.2$, $\lambda_a = 20.0$, and $\sigma^2=0.5$. For experiments involving pairwise alignment losses, $\gamma_1 = 0.2$ and $\gamma_2=0.5$ have the best performance on the validation dataset. We use \texttt{ADAM} as the optimizer with learning rate $10^{-4}$ over 500 epochs. Batch size is 2. We use MSE as the similarity loss.\\
{\bf \emph{Aladdin}}: When specifying the atlas as in Eq.~\eqref{eq:forward_opt_solution}, the following hyperparameters achieve the best performance in the validation dataset: the coefficient for the similarity loss is $10.0$, $\lambda = 1000.0$, $\gamma_1 = 1.0$ and $\gamma_2 = 5.0$. We use \texttt{ADAM} as the optimizer to learn network parameters with learning rate $10^{-4}$ over 500 epochs. Batch size is 2. We use MSE and NCC as the similarity losses in different experiments. When learning the atlas, we use another \texttt{SGD} optimizer with learning rate $10^4$. When using NCC as the similarity loss, the coefficient of the similarity loss changes to $0.3$ and $\gamma_2 = 0.15$.

\subsection{Inverse deformation map calculation}
For our approach and Dalca~\etal~\cite{dalca2019learning}, the inverse transformations are directly available. For all other baseline models, where inverse deformation maps are not available as part of the implementations, we obtain them numerically by solving $\argmin_{\phi}\|\Phi \circ \phi - \text{Id}\|_2^2 + \|\phi \circ \Phi - \text{Id}\|_2^2$, where $\Phi$ is the known deformation map and $\phi$ is the sought-for inverse transformation map. We optimize using \texttt{ADAM}~\cite{kingma2014adam} with learning rate $10^{-4}$ until convergence. The loss is defined as $\mathcal{L} = \|\Phi \circ \phi - \text{Id}\|_2^2 + \|\phi \circ \Phi - \text{Id}\|_2^2$ and $\phi$ is the to-be-optimized parameter.

\subsection{Example of an atlas built using \textbf{\emph{Aladdin}}}
Fig.~\ref{fig:atlas_comparison} shows that images differ a lot in the image population. Hence the initial averaged atlas without affine pre-alignment is very fuzzy. Note that the averaging equation in the middle row is Eq.~\eqref{eq:backward_opt_solution} and the averaging equation in the bottom row is Eq.~\eqref{eq:forward_opt_solution} due to their \emph{backward} and \emph{forward} nature. After the affine alignments, the averaged atlas is clearer and more anatomical structures can be observed. \emph{Aladdin} results in an even clearer structure, which means training images align well in atlas space. Besides, \emph{Aladdin} with pairwise alignment loss results in slightly better alignment for the training images as indicated by the slightly clearer anatomical structures in the bottom row.  

\subsection{Evaluation measures for atlas building and registration in atlas space}
In this work, we define evaluation measures in atlas space (i.e., $d^{atlas}_{k}$) as 
\begin{equation}
\label{eq:evaluate_in_atlas_space_consensus}
    d^{atlas}_k = \frac{1}{M}\sum_{i=1}^M \text{Dice}\big(S^k_i \circ \Phi_i \,, \texttt{V}(\{S^k_i \circ \Phi_i\}_{i=1...M})\big)\,,
\end{equation}
in Sec.~\ref{sec:performance_and_analysis}. In this definition, we first use a plurality voting scheme to obtain the consensus segmentation among all warped segmentations and then compare it with each warped segmentation. Note that in some other work~\cite{he2020unsupervised,he2021learning} evaluation measures in atlas space are defined without a consensus segmentation, i.e.,
\begin{equation}
\label{eq:evaluate_in_atlas_space_pairwise}
    \tilde{d}^{atlas}_k = \frac{2}{M(M-1)}\sum_{(i,j) \in \Gamma} \text{Dice}\big(S^k_i \circ \Phi_i \,, S^k_j \circ \Phi_j\big)\,,
\end{equation}
where $\Gamma = \{ (i, j) | i=1, 2, ..., M, j=1, 2, ..., M, i < j  \}$ is the set of all pairwise index combinations where the first index is smaller than the second index. Both evaluation measures are reasonable but evaluating Eq.~\eqref{eq:evaluate_in_atlas_space_consensus} has $O(N)$ complexity while evaluating Eq.~\eqref{eq:evaluate_in_atlas_space_pairwise} has $O(N^2)$ complexity. Therefore, we choose Eq.~\eqref{eq:evaluate_in_atlas_space_consensus} as the evaluation measure in atlas space to save computational time especially for images with multiple structures. The same reasoning also applies to our atlas-as-a-bridge evaluation measure $d^{bridge}_k$ in Sec.~\ref{sec:performance_and_analysis}.

\subsection{Regularization}
\label{app:regularization_more_explanation}
\emph{Aladdin} is the only atlas building approach that uses the bending energy for regularization. The benefit of using the bending energy comes from the fact that it is invariant to affine transformations. Therefore no separate affine pre-registration is required. Assume we have an affine transformation of the form $\Phi^{aff} = Ax + x + b$, parameterized s.t. for $A = 0$, $b = 0$ we obtain the identity transform. Therefore, we have $\frac{\partial{\Phi^{aff}}}{\partial x} = A + \text{Id}$ and $\frac{\partial^2{\Phi^{aff}}}{\partial x^2} = 0$. Hence, the bending energy regularization zeros out any affine contributions. Consequently, in contrast to approaches with zero or first order derivative terms in their regularizer, \emph{Aladdin} can \emph{simultaneously} capture affine and nonparametric deformations.

\end{document}